\documentclass[pra,aps,preprint,superscriptaddress,showpacs,nofootinbib]{revtex4-1}

\usepackage[dvips]{graphicx}
\usepackage{amsmath,amssymb,amsthm,mathrsfs,amsfonts,dsfont}
\usepackage{braket}
\usepackage{bm}
\usepackage{enumerate}
\usepackage{multirow}

\newcommand{\ea}{\mathrm{EA}}
\newcommand{\oexp}{\omega_\mathrm{exp}}
\newcommand{\dest}{\delta_\mathrm{est}}
\usepackage{url}

\usepackage{epsfig}
\usepackage{subfigure, epsfig}
\usepackage{graphicx}
\usepackage{color}
\usepackage[10pt]{moresize}

\usepackage{amscd}

\begin{document}

\title{Device-independent quantum random number generation}
\author{Yang Liu}
\affiliation{Shanghai Branch, National Laboratory for Physical Sciences at Microscale and Department of Modern Physics, University of Science and Technology of China, Shanghai, 201315, P.~R.~China}
\affiliation{Shanghai Branch, CAS Center for Excellence and Synergetic Innovation Center in Quantum Information and Quantum Physics, University of Science and Technology of China, Shanghai 201315, P.~R.~China}

\author{Qi Zhao}
\affiliation{Center for Quantum Information, Institute for Interdisciplinary Information Sciences, Tsinghua University, Beijing 100084, P.~R.~China}

\author{Ming-Han Li}
\author{Jian-Yu Guan}
\affiliation{Shanghai Branch, National Laboratory for Physical Sciences at Microscale and Department of Modern Physics, University of Science and Technology of China, Shanghai, 201315, P.~R.~China}
\affiliation{Shanghai Branch, CAS Center for Excellence and Synergetic Innovation Center in Quantum Information and Quantum Physics, University of Science and Technology of China, Shanghai 201315, P.~R.~China}

\author{Yanbao Zhang}
\affiliation{NTT Basic Research Laboratories and NTT Research Center for Theoretical Quantum Physics, NTT Corporation, 3-1 Morinosato-Wakamiya, Atsugi, Kanagawa 243-0198, Japan}

\author{Bing Bai}
\affiliation{Shanghai Branch, National Laboratory for Physical Sciences at Microscale and Department of Modern Physics, University of Science and Technology of China, Shanghai, 201315, P.~R.~China}
\affiliation{Shanghai Branch, CAS Center for Excellence and Synergetic Innovation Center in Quantum Information and Quantum Physics, University of Science and Technology of China, Shanghai 201315, P.~R.~China}

\author{Weijun Zhang}
\affiliation{State Key Laboratory of Functional Materials for Informatics, Shanghai Institute of Microsystem and Information Technology, Chinese Academy of Sciences, Shanghai 200050, P.~R.~China}

\author{Wen-Zhao Liu}
\author{Cheng Wu}
\affiliation{Shanghai Branch, National Laboratory for Physical Sciences at Microscale and Department of Modern Physics, University of Science and Technology of China, Shanghai, 201315, P.~R.~China}
\affiliation{Shanghai Branch, CAS Center for Excellence and Synergetic Innovation Center in Quantum Information and Quantum Physics, University of Science and Technology of China, Shanghai 201315, P.~R.~China}

\author{Xiao Yuan}
\affiliation{Shanghai Branch, National Laboratory for Physical Sciences at Microscale and Department of Modern Physics, University of Science and Technology of China, Shanghai, 201315, P.~R.~China}
\affiliation{Shanghai Branch, CAS Center for Excellence and Synergetic Innovation Center in Quantum Information and Quantum Physics, University of Science and Technology of China, Shanghai 201315, P.~R.~China}
\affiliation{Center for Quantum Information, Institute for Interdisciplinary Information Sciences, Tsinghua University, Beijing 100084, P.~R.~China}

\author{Hao Li}
\affiliation{State Key Laboratory of Functional Materials for Informatics, Shanghai Institute of Microsystem and Information Technology, Chinese Academy of Sciences, Shanghai 200050, P.~R.~China}

\author{W. J. Munro}
\affiliation{NTT Basic Research Laboratories and NTT Research Center for Theoretical Quantum Physics, NTT Corporation, 3-1 Morinosato-Wakamiya, Atsugi, Kanagawa 243-0198, Japan}

\author{Zhen Wang}
\author{Lixing You}
\affiliation{State Key Laboratory of Functional Materials for Informatics, Shanghai Institute of Microsystem and Information Technology, Chinese Academy of Sciences, Shanghai 200050, P.~R.~China}

\author{Jun Zhang}
\affiliation{Shanghai Branch, National Laboratory for Physical Sciences at Microscale and Department of Modern Physics, University of Science and Technology of China, Shanghai, 201315, P.~R.~China}
\affiliation{Shanghai Branch, CAS Center for Excellence and Synergetic Innovation Center in Quantum Information and Quantum Physics, University of Science and Technology of China, Shanghai 201315, P.~R.~China}

\author{Xiongfeng Ma}
\affiliation{Center for Quantum Information, Institute for Interdisciplinary Information Sciences, Tsinghua University, Beijing 100084, P.~R.~China}

\author{Jingyun Fan}
\author{Qiang Zhang}
\author{Jian-Wei Pan}
\affiliation{Shanghai Branch, National Laboratory for Physical Sciences at Microscale and Department of Modern Physics, University of Science and Technology of China, Shanghai, 201315, P.~R.~China}
\affiliation{Shanghai Branch, CAS Center for Excellence and Synergetic Innovation Center in Quantum Information and Quantum Physics, University of Science and Technology of China, Shanghai 201315, P.~R.~China}

\begin{abstract}
Randomness is critical for many information processing applications, including numerical modeling and cryptography \cite{shannon1949communication, metropolis1949monte}. Device-independent quantum random number generation \cite{Mayers98} (DIQRNG)  based on the loophole free violation of Bell inequality \cite{Hensen_Loophole_2015, Shalm15, Giustina15, Rosenfeld17} produces unpredictable genuine randomness without any device assumption and is therefore an ultimate goal in the field of quantum information science \cite{ma2016quantum, RevModPhys.89.015004, Acin_Certified_2016}. However, due to formidable technical challenges, there were very few reported experimental studies of DIQRNG \cite{pironio2010, Christensen13, Bierhorst17, Liu_High_2018}, which were vulnerable to the adversaries. Here we present a fully functional DIQRNG against the most general quantum adversaries \cite{miller2017universal, vazirani2011certifiable,arnon2016simple}. We construct a robust experimental platform that realizes Bell inequality violation with entangled photons with detection and locality loopholes closed simultaneously. This platform enables a continuous recording of a large volume of data sufficient for security analysis against the general quantum side information and without assuming independent and identical distribution. Lastly, by developing a large Toeplitz matrix (137.90 Gb $\times$ 62.469 Mb) hashing technique, we demonstrate that this DIQRNG generates $6.2469\times 10^7$ quantum-certified random bits in 96 hours (or 181 bits/s) with uniformity within $10^{-5}$. We anticipate this DIQRNG may have profound impact on the research of quantum randomness and information-secured applications.
\end{abstract}

\maketitle

Creating a full DIQRNG based on the violation of Bell inequality necessitates to fulfill two sets of conditions rigorously and simultaneously. First, one must close both detection and locality loopholes simultaneously in the Bell test experiment\cite{Hensen_Loophole_2015, Shalm15, Giustina15, Rosenfeld17}. While recent remarkable progress in the loophole free test of Bell inequality provides a favorable condition to realize a full DIQRNG, the implementation demands unprecedented detection efficiency and system stability and therefore still remains a formidable challenge. Previous experimental studies of DIQRNG based on Bell test experiment only closed the detection loophole\cite{pironio2010, Christensen13, Liu_High_2018}. Second, one must not assume independent and identical distribution (i.i.d.) and must consider both classical and quantum side information in the security analysis, and the production of random bits must occur at a non-vanishing rate and be noise-tolerant. In the i.i.d. scenario, because Alice and Bob perform the same measurement on the same quantum state in each experimental trial, the security analysis can be greatly simplied by examining a single experimental trial. However, the i.i.d assumption generally fails due to issues such as memory effect and time-dependent behaviour. Although the security analysis against the most general quantum adversaries without i.i.d assumption was rigorously proven\cite{Vazirani14, Miller14, Chung14, coudron2014infinite, dupuis2016entropy, arnon2016simple}, a security analysis method efficient enough for finite-data size came into being only very recently. Exploiting the entropy accumulation theorem\cite{dupuis2016entropy}, Arnon-Friedman, Renner and Vidick recently proposed a DIQRNG analysis method that does not use the i.i.d assumption and considers the quantum side information, which nevertheless produces randomness with yield approaching the value for the i.i.d. case\cite{arnon2016simple}. Previous experiments\cite{pironio2010, Christensen13} only considered classical side information, and their security against the most general quantum adversaries is unknown. Here we report in this Letter the creation of the full DIQRNG by rigorously satisfying the two sets of conditions in experiment and security analysis.  We note that another work studying DIQRNG in parallel to our work also closed both detection and locality loopholes in the experiment\cite{Bierhorst17}, which however only considered the classical side information. Our experiment produces $6.2469\times 10^7$ genuinely quantum-certified random bits in 96 experimental hours (or 181 bits/s), with total failure probability less than $10^{-5}$. This full DIQRNG may open a new avenue to a wide range of applications including the study of fundamental physics. 

Our realization of DIQRNG is based on a sequence of Bell test experiments in the format of the Clauser-Horne-Shimony-Holt game\cite{CHSH}. Neither modeling of the physical apparatus nor relation between different experimental trials is assumed. Time-dependent or memory-like effect may happen across experimental trials. In experimental trial $i$, Alice and Bob who are spatially separated each receives a photon of an entangled pair. Alice (Bob) receives a bit from a quantum random number generator, $x_i$ ($y_i$) $\in\{0,1\}$, as input to set her (his) measurement base choice, which is not affected by the base choice and measurement outcome of Bob (Alice). Their choices of measurement settings are not affected by the emission event of the entangled photon pairs, vice versa. In other words, these events are separated spacelike to satisfy the no-signaling condition. We assume the two random inputs, $x_i$ and $y_i$ are created independently and freely irrelevant to the rest parts of the experiment, and the creation follows i.i.d for all of the $n$ trials. The corresponding measurement outcome is $a_i$ ($b_i$) $\in\{0,1\}$. We assign a CHSH game value $J_i =1$ if $ a_i\oplus b_i=x_i\cdot y_i $ and $0$, otherwise.

We consider the case of uniform input for all of the $n$-experimental trials, $p(xy)=1/4$, and leave the discussion for more general settings in Supplementary Information I.A. We obtain the CHSH game value $\bar{J}$ for all $n$ experimental trials as
\begin{equation}
\label{Eq:average}
	\bar{J} = \frac{1}{n}\sum_{j=1}^{n}J_i-3/4.
\end{equation}

The experiment is subject to various loss mechanisms. We require that the photon loss must be low enough for the experiment to be free of detection loophole. Any strategies based on local hidden variable models bound $\bar{J}\le0$. Therefore, $\bar{J} > 0$, indicates that the outcomes cannot be pre-determined and hence the presence of unpredictable quantum randomness in the outcomes.

The amount of unpredictable randomness that can be extracted against quantum side information $E$ is quantified by the smooth min-entropy, $H_{\min}^{\varepsilon_s}(\textbf{AB}|\textbf{XY}E)$ \cite{arnon2016simple}, which is bounded by,
\begin{equation}
\label{Eq:randomness}
	H_{\min}^{\varepsilon_s}(\textbf{AB}|\textbf{XY}E) \ge n\cdot R_{opt}(\varepsilon_s,\varepsilon_{\ea}, \oexp).
\end{equation}
Here $\textbf{A}$ ($\textbf{B}$) and $\textbf{X}$ ($\textbf{Y}$) denote the output and input sequences of Alice (Bob), $\varepsilon_s$ the smoothing parameter, $\oexp$ the expected CHSH game value, $\varepsilon_{\ea}$ the failure probability for entropy accumulation protocol. As a conservative estimation, we take the lower bound $R_{opt}(\varepsilon_s,\varepsilon_{\ea}, \oexp)$ as the theoretical amount of randomness on average for each trial. A complete description of function $R_{opt}$ is presented in Supplementary Information I.B. Therefore, from the raw data obtained in all $n$ experiment trials, we can extract random bits created with genuine unpredictability as:  for a given failure probability of less than $2^{-t_e}$, we apply the Toeplitz-matrix hashing extractor with a matrix of size $n\times H_{\min}^{\varepsilon_s}(\textbf{AB}|\textbf{XY}E) - t_e$ to extract $H_{\min}^{\varepsilon_s}(\textbf{AB}|\textbf{XY}E) - t_e$ random bits that is $\varepsilon_s+\varepsilon_{EA}+2^{-t_e}$ close to the uniform distribution. Here, we set $t_e = 100$.

\begin{figure*}[tbh]
\centering
\resizebox{11cm}{!}{\includegraphics{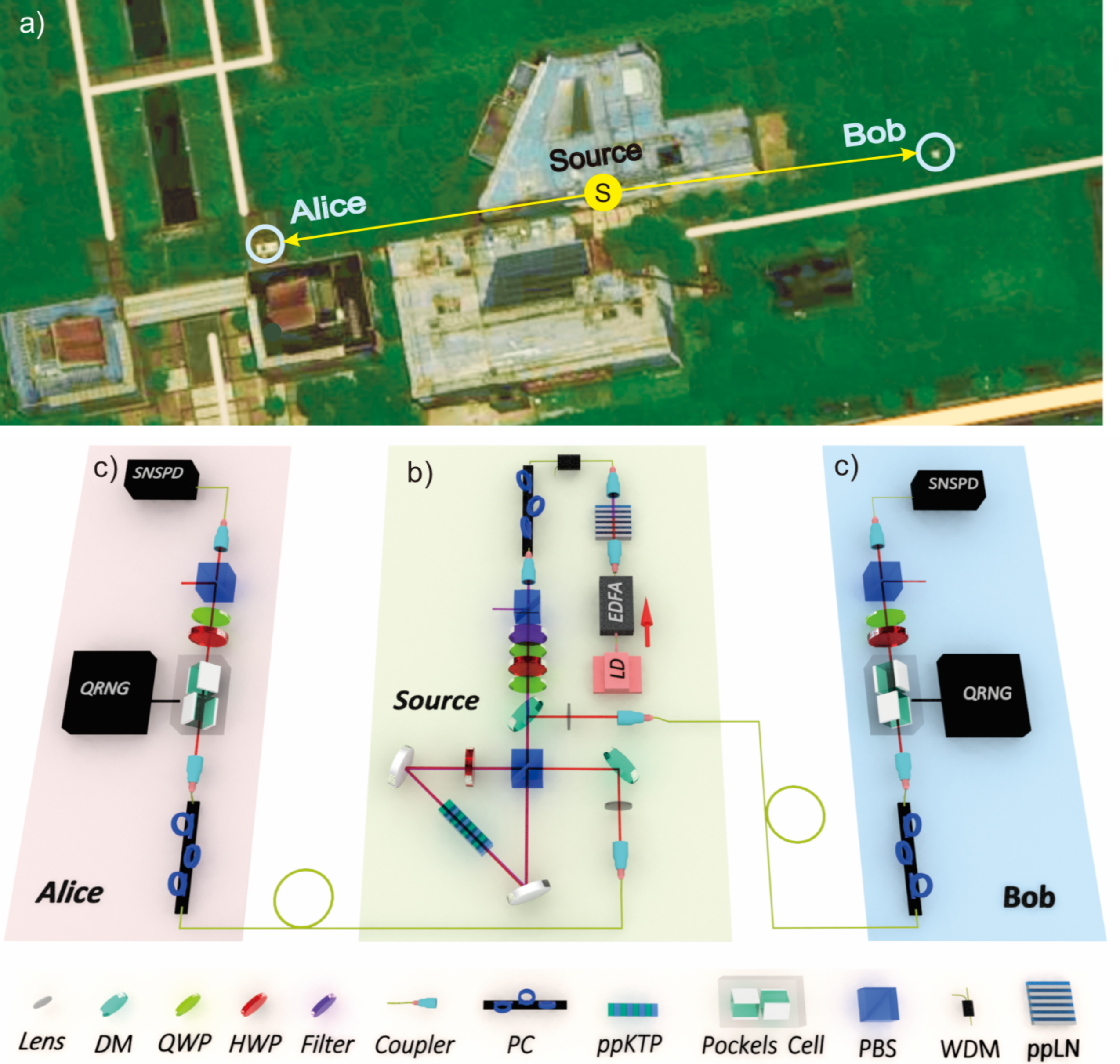}}
\caption{Schematics of the experiment.
a) A bird's-eye view  of the experimental layout. Alice's and Bob's measurement stations are on the opposite sides of the emission source of entangled photon pairs, with direct free space distance with respect to the source measured to be 93$\pm$2 m and 90$\pm$2 m, respectively. (to be continued in the next page.)
}
\label{Fig:DISetup}
\end{figure*}
\addtocounter{figure}{-1}

\begin{figure*}[tbh]
\centering
\caption{
b) Creation of pairs of entangled photons: Light pulses of 10 ns, 200 kHz from a 1560 nm seed laser (LD) are amplified by an erbium-doped fiber amplifier (EDFA), and frequency-doubled in an in-line periodically poled lithium niobate (PPLN) waveguide. With the residual 1560 nm light removed by a wavelength-division multiplexer (WDM) and spectral filters, the 780 nm light pulses are focused into a periodically poled potassium titanyl phosphate (PPKTP) crystal in a Sagnac loop to generate polarization entangled photon pairs. A set of quarter-wave plate (QWP) and half-wave plate (HWP) are used to control the relative amplitude and phase in the created polarization-entangled two-photon state. The residual 780 nm pump light is removed by dichroic mirrors (DMs) in both the source and the measurement stations. The two photons of an entangled pair at 1560 nm travel in opposite directions to two remote measurement stations, where they are subject to polarization state measurements.
c) Single photon polarization measurement: The single photons exit the fibre, complete the polarization state measurement in free space, and are collected into single mode optical fibres to be detected by superconducting nanowire single-photon detectors (SNSPD). The apparatus to perform single-photon polarization measurement consists of a Pockcels cell, a QWP, a HWP and a PBS.  Quantum random number generators are used to trigger the Pockels cell to switch between two polarization orientations that are determined to have a maximal violation of the Bell inequality as described in the main text.  A time-digital convertor (TDC) is used to time-tag the events for random number generation and single-photon detection.
}
\label{Fig:DISetupComments}
\end{figure*}

Fig.~\ref{Fig:DISetup} is the experimental schematics for the creation and detection of entangled photon pairs at low loss. We choose to create entangled photon pairs at 1560 nm based on spontaneous parametric downconversion, which has negligible loss propagating through 100 meter optical fibre. We enclose a periodically poled potassium titanyl phosphate (PPKTP) crystal with poling period 46.5 $\mu m$ in a Sagnac loop. With the injection of pump pulses at wavelength of 780 nm and pulse width of 10 ns at a repetition rate of 200 kHz, the loop emits polarization-entangled photon pairs at 1560 nm. We set the beam waist to be 180 $\mu m$ for the pump beam (780 nm) and 85 $\mu m$ for the collection beam (1560 nm) to optimize the efficiency to couple the created photons at 1560 nm into optical fibre. We place all elements for generating and collecting entangled photon pairs into single mode optical fibre on a (1 m $\times$ 1 m) breadboard to improve the system stability, with ambient temperature stabilized to be within $\pm 1^oC$.

\begin{figure}[tbh]
\centering
\resizebox{8cm}{!}{\includegraphics{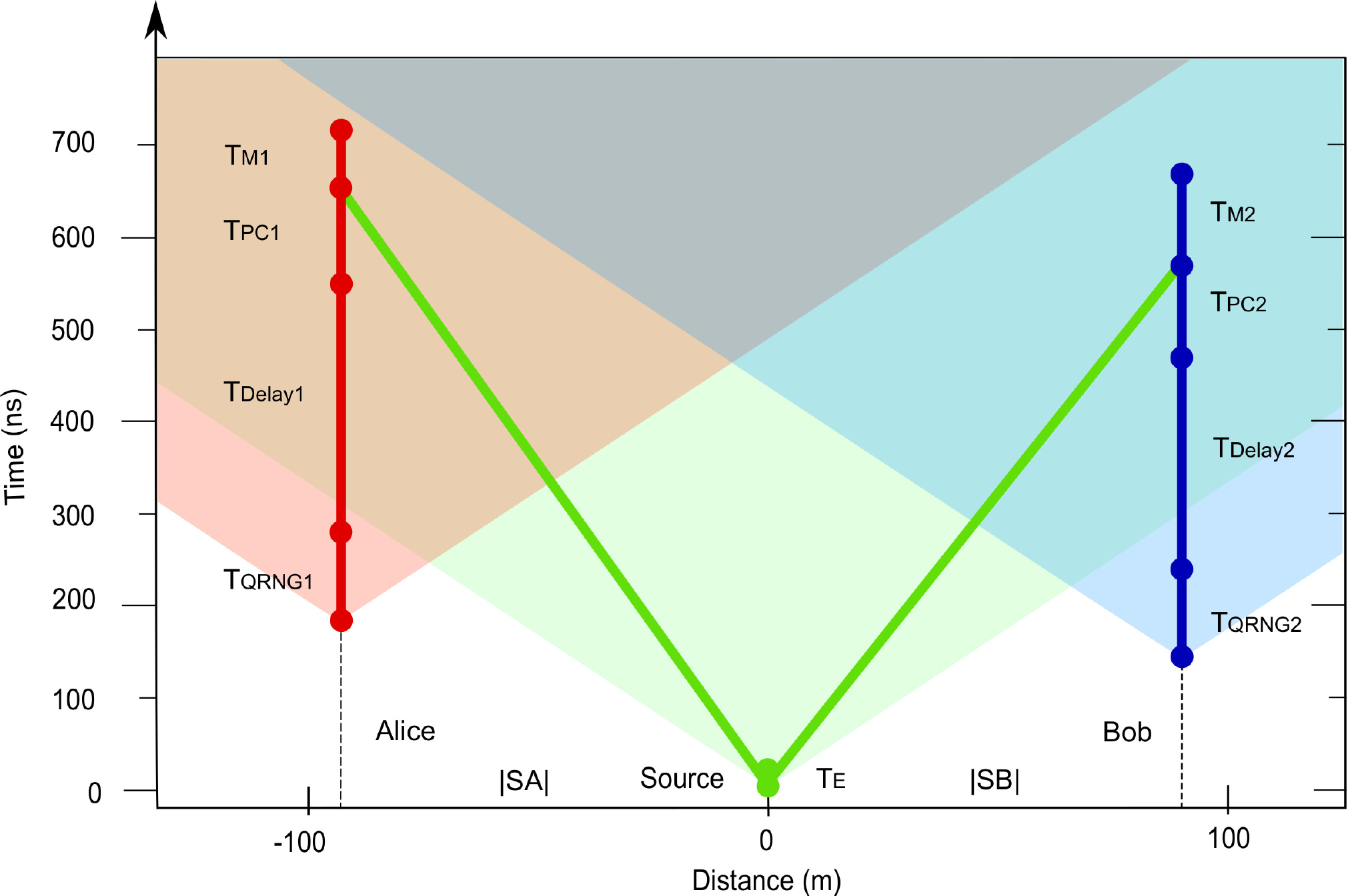}}
\caption{Spacetime diagram for the experimental design.
$T_E=10$ ns is the time elapse to generate a pair of entangled photons. $T_{QRNG1,2}$ is the time elapse to generate random bits to switch the Pockels cell. $T_{Delay1,2}$ is the time elapse for QRNG to delay delivering random bits to the Pockcels cell. $T_{PC1,2}$ is the time elapse for the Pockcels cell to be ready to perform state measurements after receiving the random bits from the QRNG. $T_{M1,2}$ is the time elapse for the SNSPD to output an electronic signal. For $T_{QRNG1}=T_{QRNG2}= 96$ ns, $T_{Delay1}=270$ ns and $T_{Delay2}=230$ ns, $T_{PC1}=112$ ns and $T_{PC2}=100$ ns, $T_{M1}= 50$ ns and $T_{M2}= 100$ ns, we place Alice's measurement station and Bob's measurement station on the opposite side of the source and 93 (90) meter from the source, and set the effective optical length between Alice's (Bob's) station and the source to be 132 m (119 m). This arrangement ensures spacelike separation between measurement event and distant base setting event and between base setting event and photon pair emission event.
}
\label{Fig:SpaceTime}
\end{figure}

We obtain an overall efficiency from creation to detection of single photons to be $(78.8\pm1.9)\%$ for Alice and $(78.5\pm1.5)\%$ for Bob \cite{Pereira2013}, the highest to-date for entangled photons, surpassing the threshold to close the detection loophole. The loss is mainly due to the limited efficiency, $94\%$, in collecting the created photon pairs into single mode optical fibre  and the limited efficiency, $\approx 92\%$, of superconducting nanowire single-photon detectors\cite{zhang2017nbn} (SNSPD) (Supplementary Information II.E), which can be technically reduced in future works. To ensure the observation of no-signaling condition, we send the two photons of a pair in opposite direction to Alice's (Bob's) measurement station, which is 93 meters (90 meters) away from the source, see Fig.~\ref{Fig:SpaceTime} for spacetime analysis (Supplementary Information II.G).

For the maximum violation of Bell inequality \cite{Eberhard93}, we create nonmaximally polarization-entangled two-photon state, $\cos(22.05^\circ)\ket{HV}+\sin(22.05^\circ)\ket{VH}$; in single photon polarization state measurements, Alice's Pockcels cell switches between two settings, either position $X_0=-83.5^\circ$ (for $x=0$) or $X_1=-119.4^\circ$ (for $x=1$), and Bob's Pockcels cell switch between position $Y_0=6.5^\circ$ (for $y=0$) or $Y_1=-29.4^\circ$ (for $y=1$), instructed by her (his) quantum random number generator. The two quantum random number generators are created based on vacuum noise fluctuation.

\begin{figure}[hbt]
  \centering
  \resizebox{8cm}{!}{\includegraphics[scale=1]{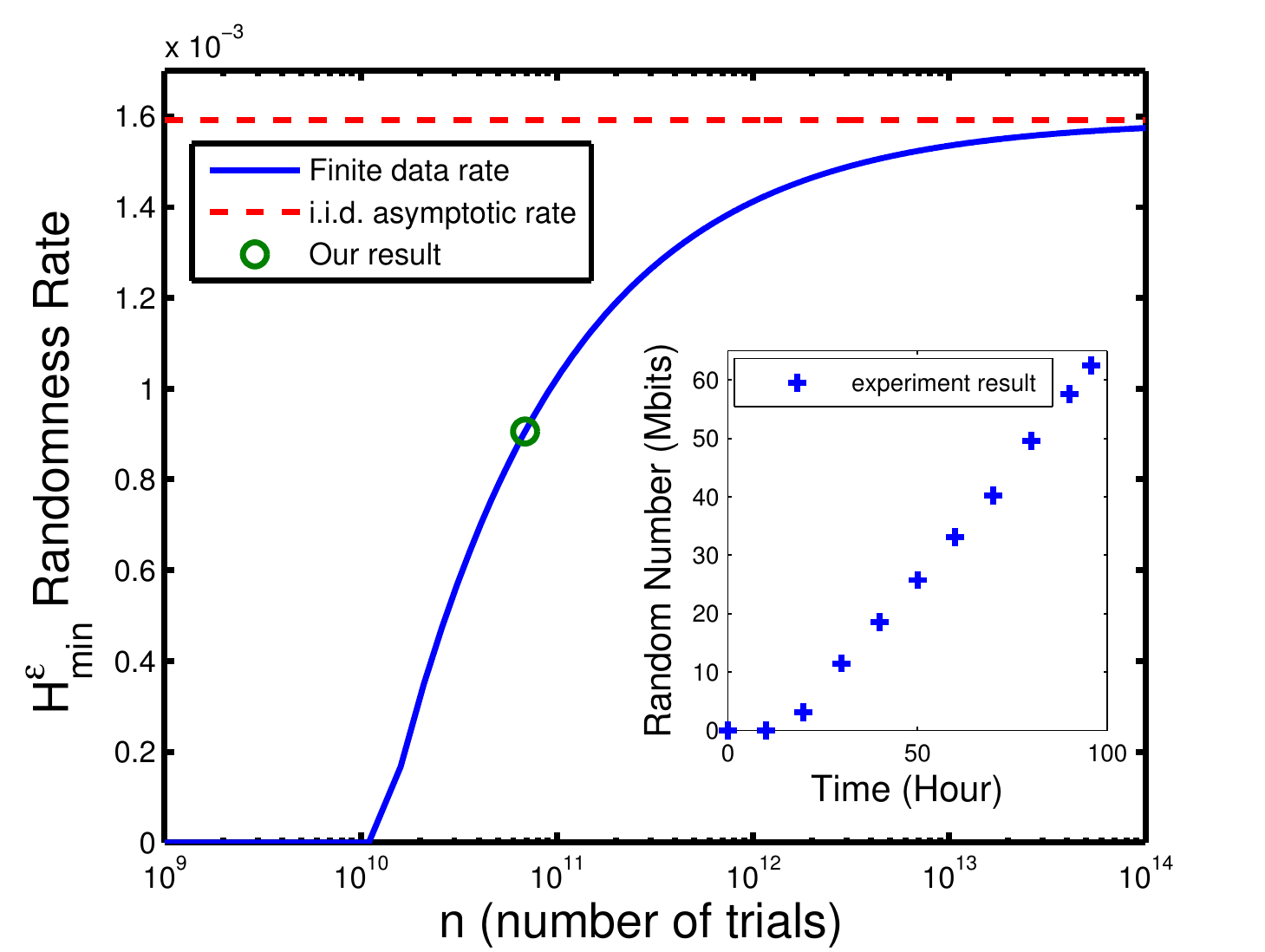}}\\
  \caption{Randomness generation versus number of experimental trials. We set the expected CHSH game value to be $\oexp = 2.757\times10^{-4}$, $\varepsilon_s = \varepsilon_{\ea} = 1/\sqrt{n}$ and $\dest = \sqrt{10/n}$ for finite data rate. }\label{Fig:randomN}
\end{figure}

Our system is robust against noise, which allows us to complete $n=6.895\times10^{10}$ experimental trials in 95.77 experimental hours without break. The predication-based-ratio analysis\cite{Zhang2011} produces an extremely small p-value $p_n= 10^{-204792}$ in the hypothesis test of local realism, indicating a strong rejection of local hidden variable models, and produces a p-value $p_n= 1$ in the hypothesis test of no signaling, showing no evidence of anomalous signaling in the experiment (Supplementary Information III.B). The CHSH game $J$-value for $n$-experimental trials is computed to be $\bar{J}=2.757 \times 10^{-4}$. By setting the expected CHSH game value to the one measured in the experiment, $\oexp = 2.757\times10^{-4}$, $\varepsilon_s = \varepsilon_{\ea} = \sqrt{1/n}=3.8 \times 10^{-6}$ and the width of the statistical confidence interval for the Bell violation estimation test $\dest = \sqrt{10/n}= 1.2042 \times 10^{-5}$, the total failure probability $\varepsilon_s+\varepsilon_{\ea}+ 2^{-t_e}<1\times 10^{-5}$. After developing a new computing technique allowing us to apply an 137.90 Gb $\times$ 62.469 Mb Toeplitz matrix hashing, we obtain $6.2469 \times 10^{7}$ genuinely quantum-certified random bits, or 181.2 bits/s, with uniformity within $ 10^{-5}$. The stream of random bits pass the NIST statistic test suite (Supplementary Information III.A). We plot the amount of randomness that can be produced by our experiment as a function of the number of experimental trials, which asymptotically approaches the optimal asymptotic value for i.i.d. as shown in Fig.~\ref{Fig:randomN}. The amount of randomness obtained in the current experiment is about 56.9\% of the optimal asymptotic value. We plot the randomness production as a function of time in the inset, demonstrating the system robustness.

In conclusion, we report the full realization of DIQRNG, which rejects local hidden variable models, is against both quantum and classical adversaries, is regardless of time-dependent behavior, outputs genuine random bits with unpredictability at high rates and is robust against noise. We anticipate that this robust DIQRNG shall stimulate more aggressive advancement in the study of randomness, for example in randomness expansion and randomness amplification\cite{kessler2017device}, and practical applications such as randomness beacon. We also anticipate this DIQRNG may help to answer more fundamental questions such as what is minimum assumption necessary for randomness generation.

\emph{Acknowledgement.}---The authors would like to thank S.-R. Zhao, Y.-H. Li, L.-K. Chen, R. Jin for experimental assistance, J. Zhong, S.-C. Shi for low temperature system maintenance, T. Peng, Y. Cao, C.-Z. Peng, Y.-A. Chen for enlightening discussions. This work has been supported by the National Key R\&D Program of China (2017YFA0303900, 2017YFA0304000), the National Natural Science Foundation of China, and the Chinese Academy of Sciences.

\section*{Supplemental Material}
\section{Theory of device-independent quantum random number generation}

Here, we generate the random number in a device-independent manner. To be more specific, our random generation is based on the violation of loophole-free Bell inequality. In this section we introduce the protocol and give randomness generation analysis.

\subsection{Protocol description}

We perform the extraction on output string $AB$ and obtain a classical string $Z$, which is close to uniform distribution even conditioned on random input $XY$ and a potential adversary's system $E$. We require the performance of the protocol in two different aspects. On one hand, for an untrusted device, the protocol is either aborted with probability less than $P_{abort}$, or the returned string $Z$ should satisfy
 \begin{equation}
    (1-P_{abort})\|\rho_{ZXYE}-\rho_{U_m}\otimes \rho_{XY}\otimes \rho_{E}  \|\le \varepsilon^{s}_{QRNG}.
    \end{equation}
Here, $\rho_{ZXYE}$ denotes the state of the classical string $Z$, the input randomness source $XY$ and the system of Eve $E$; $\rho_{U_m}$ denotes a string of uniform distribution random bits. This performance is quantified by soundness error $\varepsilon^{s}_{QRNG}$. On the other hand, we also require the honest device does not abort with probability greater than $1-\varepsilon^{c}_{QRNG}$, where $\varepsilon^{c}_{QRNG}$ is the completeness error.

For the Bell test experiment, the input are required to be random, which consumes random numbers, and in our case for each trial of Bell test, 2 random bits are consumed. In order to obtain the randomness expansion, the spot-checking protocol \cite{Coudron2013, miller2017universal, arnon2016simple}, where Bell test is only run with probability $q$ (usually very small) and the input is fixed with probability ($1-q$).

A CHSH spot-checking protocol is shown as follows.
\begin{enumerate}
	\item Bell test:
\begin{enumerate}
\item
A classical bit string $\mathbf{t}=(t_1,\cdots,t_n)$ is generated, for any $i\in \{1,2,\cdots,n\}$, $ t_i\in\{0,1\}$ according to the distribution $(1-q,q)$.
\item
The game involves two players, Alice and Bob, who are restricted to communicate with each other in the game. In the $i$th trial, they choose the input random bits $x_i$ and $y_i$ according to the bit $x_i$. If $t_i=1$, this trial is a test trial, which is used to test the existence of adversary. Then Alice and Bob randomly input $x_i=\{0,1\}$, $y_i=\{0,1\}$ and are required to output bits $a_i$ and $b_i$.  If $t_i=0$, this trial is a generation trial with fixed input string, $x_i=0$ $y_i=0$
\item
We calculate the game score according to $\mathbf{t}$, $a_i$, $b_i$, $x_i$ and $y_i$. If $t_i=1$, a score is recorded according to a pay-off function $J_i(a_ib_ix_iy_i)\in \{0,1\}$, which is given as follows.
\begin{equation}
J_i(a_ib_ix_iy_i) = \begin{cases}1 &a_i+b_i = x_i*y_i \\ 0 & a_i+b_i \neq x_i*y_i\end{cases}
\end{equation}
If $t_i=0$, $J_i=0$.
\item
(b)-(c) steps are repeated in total $n$ trials.
\item
We abort the protocol when $\sum_i J_i/n < \oexp q - \dest$. Otherwise, the protocol can be used for quantum random number generation. And the completeness error is upper bounded by $\varepsilon^c_{QRNG}\le \exp{(-2n\dest^2)}$.
\end{enumerate}
\item Randomness estimation: conditioned on the Bell test is not aborted, either the protocol aborts with probability less than $\varepsilon^c_{QRNG}$ or the amount of randomness
    is given by Eq.~\eqref{Eq:rawkey}.
\item Randomness extraction: with failure probability less than $2^{-t_e}$, we can extract $n\cdot R_{opt}(\varepsilon_s,\varepsilon_{EA}) - t_e$ random bits that is
    $\varepsilon^s_{QRNG}$ close to a uniform distribution by using the Toeplitz-matrix hashing, where $\varepsilon_s$ is the smoothing parameter, $\varepsilon_{EA}$ is the error probability of the entropy accumulation protocol. 
\end{enumerate}

\subsection{Estimation of randomness production}

In this work, we first apply the security proof in Ref.~\cite{arnon2016simple}. Ref.~\cite{arnon2016simple} uses the tool, entropy accumulation theorem (EAT) \cite{dupuis2016entropy} to reduce the complex multi-trial protocol to the i.i.d. case, in other words, from coherent attack case to collective case. In Ref.~\cite{arnon2016simple}, they prove that the device independent protocol satisfies the requirements of an EAT channel. Next, the randomness production based on the violation of CHSH inequality \cite{pironio2009device} under the stronger i.i.d. assumption can be connect to the non-i.i.d. case and gives a good lower bound for the generated randomness. To be more specific, the device independent proofs against coherent attacks try to find the lower bound for $H_{\min}(\textbf{AB}|\textbf{XY}E)$ when given the Bell violation value. Ref.~\cite{arnon2016simple} provides a framework to reduce the general attack to collective attack by considering $H(AB|EXY)$ instead of $H_{\min}(AB|EXY)$. Collective attack supposes that Eve prepares the state $\rho_{ABE}^{\otimes n}$, and the measurement setting for each round is the same. So this framework greatly simplifies the security proof because each round is separately independent. 

In the Theorem 10 of \cite{arnon2016simple}, the optimal randomness yield is given by
\begin{equation}\label{Eq:rawkey}
H_{min}^{\varepsilon_s}(\textbf{AB}|\textbf{XY}E)
 \ge  n \cdot R_{opt}(\varepsilon_s, \varepsilon_{EA}).
\end{equation}
where the smoothed min-entropy $H_{min}^{\varepsilon_s}(\textbf{AB}|\textbf{XY}E)$ evaluates the amount of extractable randomness, $\varepsilon_s$ is the smoothing parameter, $\varepsilon_{EA}$ is the error probability of the entropy accumulation protocol. Here, $R_{opt}(\varepsilon_s, \varepsilon_{EA})$ is defined as
\begin{align}
g(p) = \begin{cases}1-h\left(\frac{1}{2}+\frac{1}{2}\sqrt{16\frac{p}{q}(\frac{p}{q}-1)+3}\right)& \frac{p}{q} \in [0,\frac{2+\sqrt{2}}{4}]\\ 1 & \frac{p}{q} \in
[\frac{2+\sqrt{2}}{4},1]\end{cases}
\end{align}
\begin{align}
f_{\min}(p,p_t) = \begin{cases}g(p)&p\leq p_t\\ \frac{d}{dp}g(p)|_{p_t}\cdot p + (g(p_t) - \frac{d}{dp}g(p)|_{p_t}\cdot p_t) & p > p_t\end{cases}
\end{align}
\begin{align}
&R(p, p_t, \varepsilon_s, \varepsilon_e)\\
 &= f_{\min}(p, p_t) - \frac{1}{\sqrt{n}}2(\log 13 + \frac{d}{dp}g(p)|_{p_t})
\sqrt{1-2\log (\varepsilon_s \cdot \varepsilon_e)}.
\end{align}
\begin{align}
R_{opt}(\varepsilon_s, \varepsilon_e) = \max_{\frac{3}{4}<\frac{p_t}{q}<\frac{2+\sqrt{2}}{4}} R(\omega_{exp}\cdot q - \delta_{est}, p_t, \varepsilon_s, \varepsilon_e).
\end{align}
Here $\delta_{est}\in (0,1)$ is the width of the statistical confidence interval for the Bell violation estimation test, which is bounded by the number of trials $n$ and the completeness error $\varepsilon^{c}_{QRNG}$, $\varepsilon^{c}_{QRNG}\le \exp(-2n \delta_{est}^2)$. $\omega_{exp}$ is the expected winning probability for an honest but noisy device, and is chosen according to the experimental winning scores $\sum_i J_i$ and $\omega_{exp}=\sum_i J_i/(nq) $. The total soundness error in the protocol consists three parts, the error in the EAT channel theorem $\varepsilon_{EA}$, the smooth parameter $\varepsilon_s$ and the failure probability in the randomness extraction process $2^{-t_e}$. Thus, we have $\varepsilon^{s}_{QRNG}=\varepsilon_{EA}+ \varepsilon_s +2^{-t_e}$. More details about the formula of $R_{opt}(\varepsilon_s, \varepsilon_{EA})$ is shown in \cite{arnon2016simple} and Supplemental Materials of Ref.~\cite{PhysRevLett.120.010503}.

In the above analysis, though EAT channel has been applied to solve the coherent attack case. Considering a practical case where the total trials $n$ is finite, the finite-data problem emerges. In order to generate randomness, $n$ is usually required to be large enough which brings a lot of troubles for the experiments. Thus an improved generation rate is vital and consequently reduce the required trials of experiment. There are few directions that could improve the finial generation rate. One direction is focusing on the result in  \cite{pironio2009device}, collective attack case and improve the generation rate. Following the same argument in Ref.~\cite{arnon2016simple}, the generation rate against coherent attacks can be obtained. Mathematically, the main problem in collective attack case is to solve the following optimization problem :
\begin{align}\label{objective}
f(S)=& \min_{\ket{\phi_{ABE}}, A_{|a|x}, B_{b|y}} H(AB|EXY)\\
s.t.~~~ \forall a,b,x,y ~~~&tr[A_{a|x}\otimes B_{b|y} \rho_{AB}] = \hat{p}_{ab|xy}.
\end{align}
The idea is inspired from \cite{acin2007device} and main ingredient is to reduce the high dimension state into 2-qubit bell diagonal state and then solve it numerically instead of giving an analytical result in \cite{acin2007device}. The details and the rigorous proof about the this method will be discussed more in the future work. Here we only use the results to show its potentials for reducing the required trials. Another interesting direction is to obtain the randomness based on the full statistics of measurement results instead of only the Bell violation. Note that Ref.~\cite{Knill2017} have already considered the full statistics of measurement results and can certify device-independent randomness without relying on any Bell inequality. However, it is only against classical side information, quantum side information case still remains to be solved.

\subsection{Biased random input}
Here in this section, we study about the case where the input is not uniformly distributed. Suppose that $x_i=\{0,1\}$, $y_i=\{0,1\}$ with probability $(p, 1-p)$, $p<1/2$. The results in \cite{arnon2016simple} can not be applied directly because the protocol in Ref.~\cite{arnon2016simple} the inputs are uniformly distributed in the test trials. In order to make the analysis suitable, we modify the Bell test as a spot-checking protocol.  Here we introduce two other random variables, $T^A_i=\{0,1\}$ and $T^B_i=\{0,1\}$ both with probability $(\frac{p}{1-p},\frac{1-2p}{1-p})$. Then the Bell test can be presented as follows.
\begin{enumerate}
\item
A classical bit string $\mathbf{T^A}=(T^A_1,\cdots,T^A_n)$  and $\mathbf{T^B}=(T^B_1,\cdots,T^B_n)$ is generated, for any $i\in \{1,2,\cdots,n\}$, $ T^A_i\in\{0,1\}$ and $T^B_i\in\{0,1\}$ according to the distribution $(\frac{p}{1-p},\frac{1-2p}{1-p})$.

\item
In the $i$th trial, they choose the input random bits $x_i$ and $y_i$ with probability $(p, 1-p)$. If $T^A_i=0$ and $T^B_i=0$, this trial is a test trial, which is used to test the existence of adversary. A score is recorded according to a pay-off function $J_i(a_ib_ix_iy_i)\in \{0,1\}$, which is given as follows.
\begin{equation}
J_i(a_ib_ix_iy_i) = \begin{cases}1 &a_i+b_i = x_i*y_i \\ 0 & a_i+b_i \neq x_i*y_i\end{cases}
\end{equation}
Otherwise, this trial is a generation trial, $J_i=0$.

\item
2 step is repeated in total $n$ trials.
\item
We abort the protocol when $\sum_i J_i/n < \oexp (\frac{p}{1-p})^2 - \dest$. Otherwise, the protocol can be used for quantum random number generation.
\end{enumerate}
After the modification, the corresponding violation can be directly used to calculate the randomness generation by the method in Ref.~\cite{arnon2016simple}. In the experiments, if the input is not uniformly distributed but its distribution is known, this modification can also be applied. Instead of deciding each trial is generation trial or test trial before the experiment in the above protocol, we can also use two random variable $T^A$ and $T^B$ to make this decision in the post data processing after the experiments. These two different processes are equivalent as long as the randomness of $T^A$ and $T^B$ can be guaranteed. Note that here we assume the input randomness obeys i.i.d distribution and is not controlled by Eve's system.

A remained problem here is that the input may be not perfectly random but controlled by adversary by a local hidden variable $\lambda$. The input could look perfectly random, $P(x,y)=1/4$. But it may be controlled by local hidden variable $\lambda$, $P(x,y|\lambda)\neq 1/4$. Usually, we refer this imperfect source as Santha-Vazirani source (SV source) \cite{barrett2011much}, $l< P(x,y|\lambda)<u$. The parameter $u$, $l$ quantify the imperfection of this randomness. If $u=l=1/4$, this is the perfect case. The imperfection will influence the classical bound for Bell test \cite{hall2010local,yuan2015clauser}. This imperfect source can be amplified in a device-independent protocol to generate perfect randomness, referred as device-independent randomness amplification \cite{colbeck2012free,kessler2017device,knill2017quantum}. At this point, we only focus on the perfect input case and leave the imperfection case as a future interesting direction. We can assume that the local hidden variable $\lambda$ is generated with the prepared states, independent of the input settings. The assumption of input randomness is reasonable because it has already passed the random number test and proved to be i.i.d distribution.

\section{ Experimental details}
\subsection{QRNG scheme and experimental setup}
Phase fluctuations generated from spontaneous emissions of a laser are random in nature. In this quantum random number generation (QRNG) system, random numbers are generated from measuring laser intensity via an interferometer which converts phase information into intensity information. In random number generation system, when the laser is operated at the threshold level quantum noise will be dominant. After modeling and quantifying the contribution of quantum noise, quantum random bits are finally generated.

\begin{figure}[htb]
\centering
\includegraphics[width=8.5 cm]{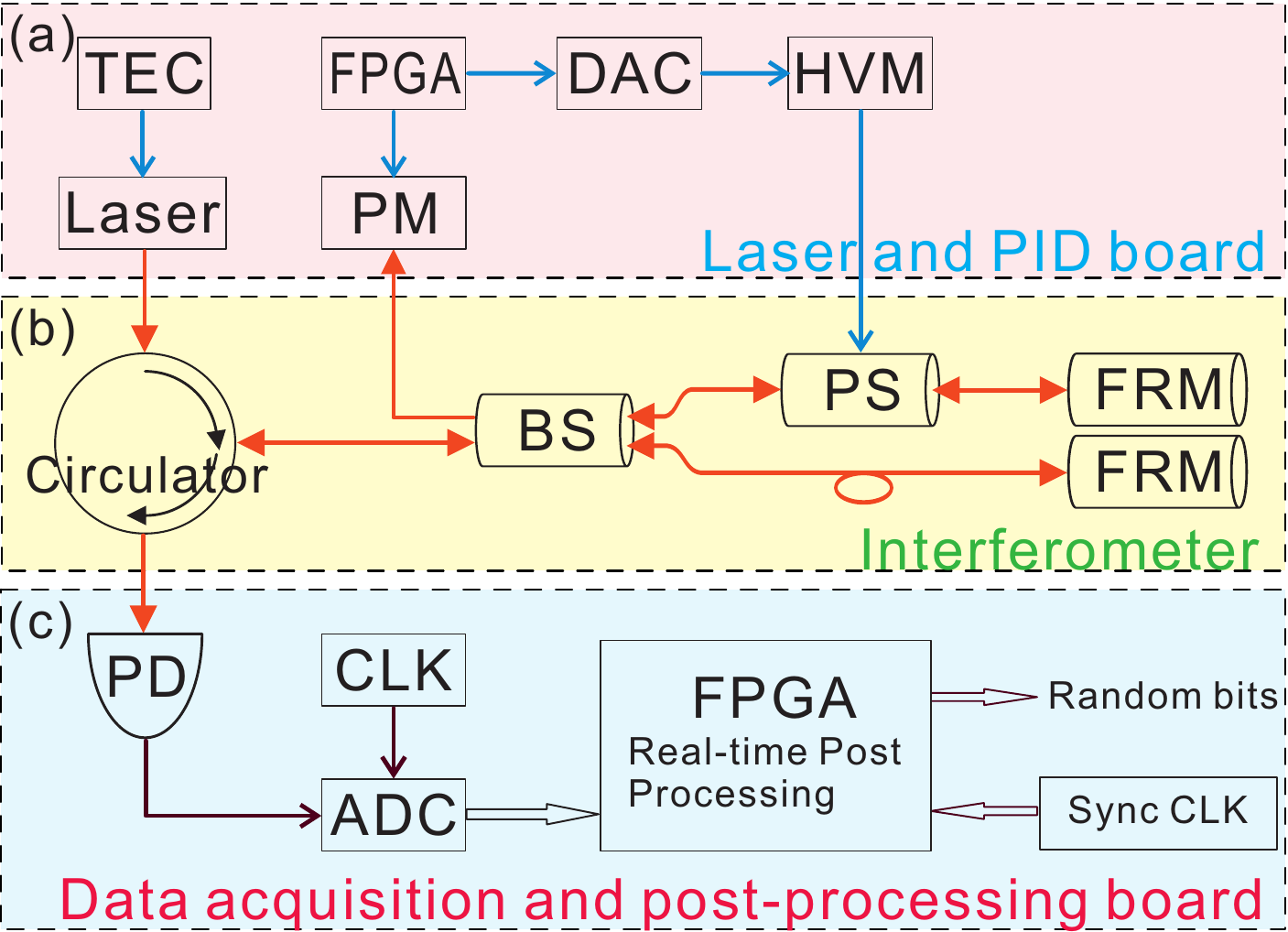}
\caption{
(a) Design diagram of the QRNG module. The key components of the QRNG module include a stable interferometer and two PCBs. One PCB includes a laser driver and a PID circuit for phase stabilization whilst the other PCB is designed for data acquisition and post-processing. The phase fluctuations signal detected by a PD is amplified and digitized by an ADC, and the sampled data are then fed into a FPGA. A real-time post-processing based on Toeplitz hashing matrix is implemented in this FPGA. The extracted random numbers are dominated by the system synchronised clock. TEC: thermoelectric cooler, LD: laser diode, FPGA: field-programmable gate array, DAC: digital-to-analog converter, HVM: high voltage module, PM: powermeter, BS: beam splitter, PS: phase shifter, FRM: Faraday rotator mirror, PD: photodetector, CLK: clock, ADC: analog-to-digital converter, Sync clock: system synchronised clock.}
\label{QRNGHardware}%
\end{figure}

The system setup is shown in Fig.~\ref{QRNGHardware}.
A $1550$ nm laser diode (LD) is driven by constant current slightly above its threshold. A thermoelectric cooler (TEC) is used for temperature stabilization. The emitted photons enter an unbalanced interferometer via a 3-port optical circulator. The interferometer consists of a circulator, a $50/50$ beam splitter (BS), two Faraday rotator mirrors (FRMs) and a Phase Shifter (PS). The FRMs can effectively remove polarization effects. The other input port of BS is monitored by an optical power meter (PM). The optical power data is transferred to a printed circuit board (PCB) which includes a field-programmable gate array (FPGA). The FPGA sends feedback data to a digital-to-analog converter (DAC), which controls a high-voltage module (HVM), by computing proportional- integral-derivative (PID) algorithm. The phase stabilization is kept at a high level through tuning the PS by HVM. The rest port of interferometer is detected by an InGaAs photodetector. The measurement results of photodetector is quantum fluctuations, which is digitized by an analog-to-digital converter (ADC).

The QRNG system is made up of two PCBs. One PCB is described in the above, which maintains the stabilization of interferometer. The other PCB is designed for data acquisition and random number signal modulating. To acquire raw random data, the ADC (AT84AD001B) samples the signal with a clock (CLK) of $1$ GSa/s and converts it to $8$ bits per sample. Then the data are fed into a FPGA, in which raw random numbers are processed to generate random numbers of better uniformity. An external synchronizing clock (sync CLK) is used to synchronize random data and output signals. Then random number data are delayed and transmitted to Modulator Drivers which drive electro-optic modulator(EOM). The other random data channel is implemented as Time-to-Digital Converter (TDC) .

\subsection{Min-entropy analysis and real-time post-processing}
Min-entropy evaluation is of vital importance to quantify the randomness of raw random data, which is acquired by ADC. We evaluate the randomness of the raw data from the QRNG following analysis in the literature \cite{Qi10,Xu12,Nie15}.

The quantum signal is given by,
\begin{equation} \label{QRNG:Icossimple}
\begin{aligned}
I(t)\propto P\sin(\Delta\theta(t))\approx P\Delta\theta(t).
\end{aligned}
\end{equation}
Therefore, the phase fluctuation of the laser source can be measured directly by the intensity of the interferometer output. In addition to the signals from phase fluctuations, the variance of the photodetector output also contains background noise. According to the theoretical model \cite{Ma13}, the quantum signal follows a Gaussian distribution. Thus, with the variance of signals from phase fluctuations, we can access the whole distribution of the quantum signal. The randomness is quantified with min-entropy, defined as follows,
\begin{equation} \label{QRNG:Hmin}
\begin{aligned}
H_\infty(X)=-\log_2(\max_{x\in\{0, 1\}^n} P_r[X=x]).
\end{aligned}
\end{equation}
Then, the min-entropy of the raw data can be evaluated via the Gaussian distribution. According to the results in literature \cite{Nie15,Zhangxg16}, the min-entropy is above $6.4$ bits per sample or $0.8$ bits per bit. The Toeplitz hashing randomness extraction is implemented in FPGA \cite{Zhangxg16,Ma13}. Firstly, we choose 1 bit from a $8$-bit ADC and perform a running parity calculation of  $x_i=(d_{4i-3}\oplus d_{4i-2}\oplus)\oplus (d_{4i-1}\oplus d_{4i})$, where $x_i$ is an element of Toeplitz matrix. Then, $16$ bit seeds are used to construct the complete Toeplitz matrix. These seeds are refreshed after every synchronizing clock circle. Finally, $16$ raw bits ($x_i,x_{i+1},...,x_{i+15}$) are multiplied to output a temporary column vector in the matrix multiplication module and extracted to single random bit. Under this circumstance, entire post-processing procedure is less than $68$ ns.

The time interval between the generation of laser phase fluctuations and the generation of extracted random bits is less than $100$ ns. Further more, several data buffer like FIFOs are used to add delay in FPGA, with which, the random bits are delayed by $270$ ns/$230$ ns for Alice/Bob before output to the Pockels cell driver.

\subsection{Preparation of optical modes for pump and collection}
Both theory \cite{Bennink2010} and experimental tests \cite{Pereira2013, Dixon2014} show that it is possible to achieve a near unity coupling efficiency in a downconversion process with appropriate focal parameters. In the experiment, for PPKTP with length to be 1 cm, we set the pump light beam waist to be 180 $\mu m$, and the collection beam waist to be 85 $\mu m$ to optimally couple daughter photons produced in the downversion into single mode fiber.

The pump light from a 780HP single mode fiber has mode field diameter of 5 $\mu$m. It is focused using a aspherical lens with f=8 mm to the PPKTP that is 70 cm away. The pump waist is measured to be 180 $\mu$m, and the beam quality $M^2$ is 1.05. For the signal (idler) collection mode, the mode field diameter of the SMF28e single mode fiber we used for collection is 10.4 $\mu$m. An aspherical lens with f=11 mm and a spherical lens with f=175 mm are used to set the beam diameter to 85 $\mu$m at the center of PPKTP crystal. The distance of the spherical lens is about 19 cm from the aspherical and about 45 cm from the PPKTP crystal.

\subsection{Determination of single photon efficiency}
We define the single photon heralding efficiency as $\eta_A=C/N_B$ and $\eta_B=C/N_A$ for Alice and Bob, in which the coincidence events (C) and the single events for Alice ($N_A$) and Bob ($N_B$) are measured in the experiment. The heralding efficiency is given by
\begin{equation} \label{Eq:heraldingEff}
\eta = \eta^{sc} \times \eta^{so} \times \eta^{fibre} \times \eta^{m} \times \eta^{det},
\end{equation}
where $\eta^{sc}$ is the efficiency to couple entangled photons into single mode optical fibre,  $\eta^{so}$ the efficiency for photons passing through the optical elements in the source, $\eta^{fibre}$ the transmittance of the fibre linking the source to the measurement station, $\eta^{m}$ the efficiency for light passing through the measurement station, and $\eta^{det}$ the single photon detector efficiency. The heralding efficiency and the transmittance of individual optical elements are listed in Table~\ref{tab:OptEffAB}, where $\eta^{fibre}$, $\eta^{m}$, $\eta^{det}$ are measured with classical light beams and NIST-traceable power meters. The coupling efficiency $\eta^{sc}$ is calculated with
\begin{equation} \label{Eq:heraldingEff}
\eta^{sc} = \frac{\eta}{\eta^{so} \times \eta^{fibre} \times \eta^{m} \times \eta^{det}}
\end{equation}

\begin{table}[htb]
\centering
  \caption{Characterization of optical efficiencies in the experiment. }
\begin{tabular}{c|c|ccccc}
\hline
 & heralding efficiency ($\eta$) & $\eta^{sc}$ & $\eta^{so}$ & $\eta^{fibre}$ & $\eta^{m}$ & $\eta^{det}$ \\
\hline
Alice  & 78.8\% & 93.9\% & \multirow{2}{*}{95.9\%} & \multirow{2}{*}{99\%} & 94.8\% & 93.2\% \\
Bob    & 78.5\% & 94.2\% &                         &                       & 95.2\% & 92.2\% \\
\hline
\end{tabular}
\label{tab:OptEffAB}
\end{table}

The transmittance of optical elements used in our experiment are listed in Table~\ref{tab:OptEff}, with which we obtain the efficiency $\eta^{so}$:
\begin{equation} \label{Eq:heraldingEff}
\eta^{so} = \eta^{AS} \times \eta^{S} \times (\eta^{DM})^4 \times \eta^{780/1560 HWP} \times \eta^{780/1560 PBS} \times \eta^{PPKTP} = 95.9\%,
\end{equation}
where we use four dichroic mirrors.

The transmittance of the 130 meter fibre between the source and the detection is $99\%$. The transmittance of the measurement station including the Pockels cell is $94.8\%$ for Alice and $95.2\%$ for Bob. The efficiency of the superconducting nanowire single-photon detector (SNSPD) \cite{zhang2017nbn} is measured to be $93.2\%$ for Alice and $92.2\%$ for Bob. The single photon heralding efficiency of the system is determined to be $\eta_A=(78.8\pm1.9)\%$ for Alice and $\eta_B=(78.5\pm1.5)\%$ for Bob with photon-counting statistic in the experiment.

\begin{table}[htb]
\centering
  \caption{The efficiencies of optical elements}
\begin{tabular}{c|c|c}
\hline
& Optical element & Efficiency\\
\hline
$\eta^{AS}$ & Aspherical lens & $99.27\%\pm0.03\%$ \\
$\eta^{S}$ & Spherical lens & $99.6\%\pm1.0\%$ \\
$\eta^{780/1560 HWP}$ & Half wave plate (780nm/1560nm) & $99.93\%\pm0.02\%$ \\
$\eta^{1560 HWP}$ & Half wave plate (1560nm) & $99.92\%\pm0.04\%$ \\
$\eta^{1560 QWP}$ & Quarter wave plate (1560nm) & $99.99\%\pm0.08\%$ \\
$\eta^{780/1560 PBS}$ & Polarizing beam splitter (780nm/1560nm) & $99.6\%\pm0.1\%$ \\
$\eta^{1560 PBS}$ & Polarizing beam splitter (1560nm) & $99.6\%\pm0.2\%$ \\
$\eta^{DM}$ & Dichroic mirror & $99.46\%\pm0.03\%$ \\
$\eta^{PPKTP}$ & PPKTP & $99.6\%\pm0.2\%$ \\
$\eta^{P}$ & Pockels cell & $98.7\%\pm0.5\%$ \\
\hline
\end{tabular}
\label{tab:OptEff}
\end{table}

\subsection{Quantum state characterization}
To maximally violate the Bell inequality in experiment, we create non-maximally entangled two-photon state $\cos(22.053^\circ)\ket{HV}+\sin(22.053^\circ)\ket{VH}$ (with $r=0.41$ for $(\ket{HV}+r\ket{VH})/\sqrt{1+r^2}$) and set the bases for single photon polarization state measurement to be  $A_1=-83.5^\circ$, $A_2=-119.38^\circ$ for Alice, $B_1=6.5^\circ$, $B_2=-29.38^\circ$ for Bob. We measure diagonal/anti-diagonal visibility in the bases set ($45^\circ, -22.053^\circ$), ($112.053^\circ, 45^\circ$) for minimum coincidence, and in the bases set ($45^\circ, 67.947^\circ$), ($22.053^\circ, 45^\circ$) for maximum coincidence, where the angles represent measurement basis  $cos(\theta)\ket{H}+sin(\theta)\ket{V}$ for Alice and Bob.

By setting the mean photon number to $\mu=0.0035$ to suppress the multi-photon effect, we measure the visibility to be $99.5\%$ and $98.5\%$ in horizontal/vertical basis and diagonal/anti-diagonal basis.

\begin{figure}[htb]
\centering
    \subfigure[]{
      \includegraphics[width=8cm]{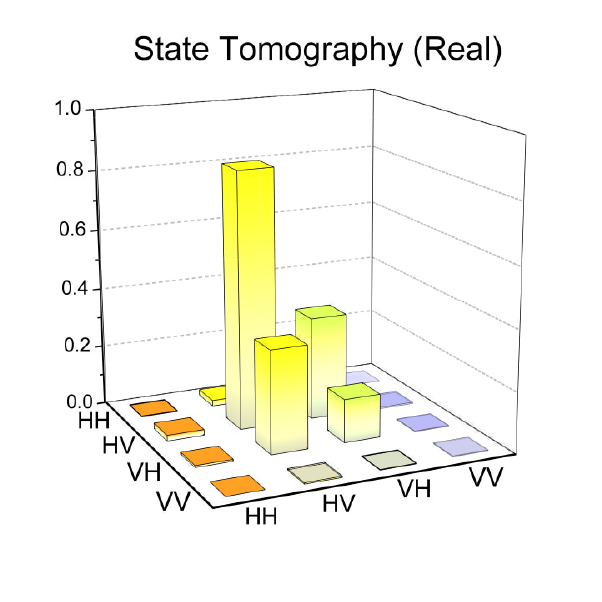}
    }
    \subfigure[]{
      \includegraphics[width=8cm]{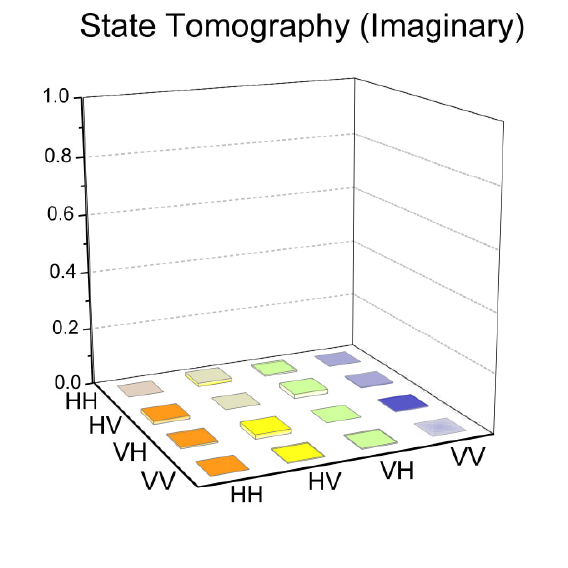}
    }
\caption{(color online)
Tomography of the produced state. The real and imaginary part are shown in (a) and (b).}
\label{Fig.Tomo}
\end{figure}

We perform state tomography on the non-maximally entangled state, the result is shown in Fig.~\ref{Fig.Tomo}, the state fidelity is $99.02\%$. We attribute the imperfection to multi-photon components, imperfect optical elements, and imperfect spatial/spectral mode matching.

\subsection{Locality and space like experimental setup}
\begin{figure}[htb]
\centering
\resizebox{13cm}{!}{\includegraphics{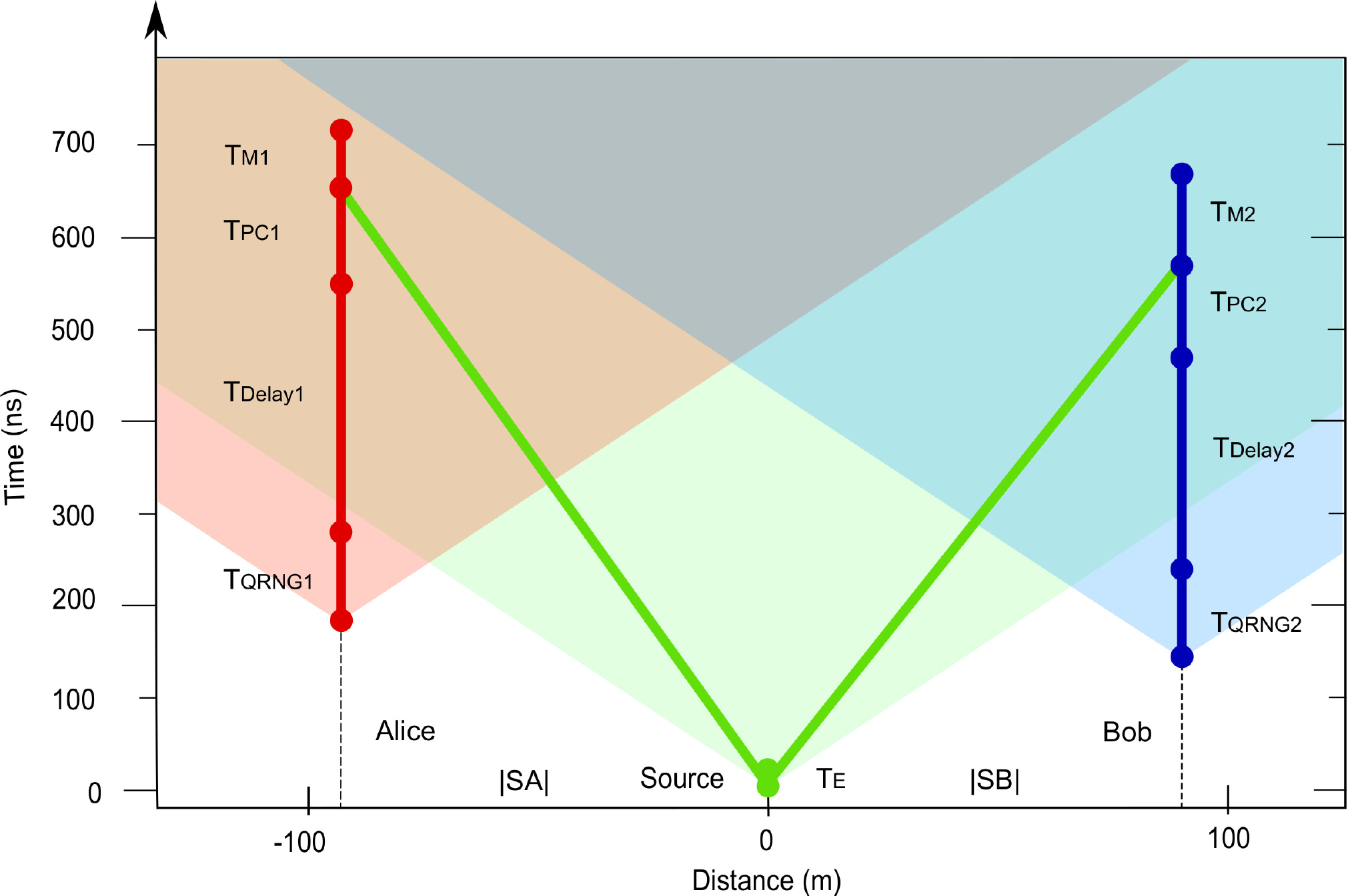}}
\caption{Space-time diagram for the experimental design. $T_E=10$ ns is the time elapse to generate a pair of entangled photons. $T_{QRNG1,2}$ is the time elapse to generate random bits to switch the Pockels cell. $T_{Delay1,2}$ is the delay between QRNG and the Pockcels cell. $T_{PC1,2}$ is the time elapse for the Pockcels cell to be ready to perform state measurements after receiving the random bits from the QRNG. $T_{M1,2}$ is the time elapse for the SNSPD to output an electronic signal. For $T_{QRNG1}=T_{QRNG2}= 96$ ns, $T_{Delay1}=270$ ns and $T_{Delay2}=230$ ns, $T_{PC1}=112$ ns and $T_{PC2}=100$ ns, $T_{M1}= 55$ ns and $T_{M2}= 100$ ns, we place Alice's measurement station and Bob's measurement station on the opposite side of the source and 93 (90) meter from the source, and set the effective optical length between Alice's (Bob's) station and the source to be 132 m (119 m). This arrangement ensures spacelike separation between measurement event and distant base setting event and between base setting event and photon pair emission event.}
\label{Fig:SpaceTimeSupp}
\end{figure}

To close the locality loophole, space-like separation must be satisfied between the state measurement events and between each measurement event and the distant setting choice event (Fig.~\ref{Fig:SpaceTimeSupp}). We can obtain

\begin{equation}
	\begin{cases}
(|SA| + |SB|) / c > T_E - (L_{SA} - L_{SB}) / c + T_{QRNG1} + T_{Delay1} + T_{PC1} +T_{M2}, \\
(|SA| + |SB|) / c > T_E + (L_{SA} - L_{SB}) / c + T_{QRNG2} + T_{Delay2} + T_{PC2} +T_{M1},
	\end{cases}
\label{Eq:SC1}
\end{equation}

where $|SA|$ = 93 m ($|SB|$ = 90 m) is the spatial distance between the entanglement source and Alice's (Bob's) measurement station, $T_E$ = 10 ns is the generation time for entangled photon pairs, which is mainly contributed by the 10 ns pump pulse duration, $L_{SA}$ = 194 m ($L_{SB}$ = 175 m) is the effective optical path which is mainly contributed by the long fiber (132 m, 119 m) between the source and Alice/Bob's measurement station, $T_{QRNG1}=T_{QRNG2}$ = 96 ns is the time elapse for quantum random number generation, $T_{Delay1}$ = 270 ns ($T_{Delay2}$ = 230 ns) is the delay between QRNG and the Pockels cells, $T_{PC1}$ = 112 ns ($T_{PC2}$ = 100 ns) including the internal delay of the Pockcels Cells (62 ns, 50 ns) and the time for the Pockcels cell to stabilize before performing single photon polarization state projection after switching which is 50 ns, $T_{M1}$ = 55 ns ($T_{M2}$ = 100 ns) is the time elapse for the SNSPD to output an electronic signal, including the delay due to fiber and cable length. 

Space-like separation must be ensured between each entangled pair creation event and the setting choice events, so we can have

\begin{equation}
	\begin{cases}
|SA| / c > L_{SA} / c  - T_{Delay1} - T_{PC1}\\
|SB| / c > L_{SB} / c  - T_{Delay2} - T_{PC2}
	\end{cases}
\label{Eq:SC2}
\end{equation}

Eq.~\ref{Eq:SC2} ensures space-like separation between the event of entangled pairs created and each event of quantum random number generation. It shows in Fig.~\ref{Fig:SpaceTimeSupp} that Alice's and Bob's quantum random number generation events are outside the future light cone (green shade) of the event.

\begin{figure}[htbp]
\centering
    \subfigure[]{
      \includegraphics[width=11cm]{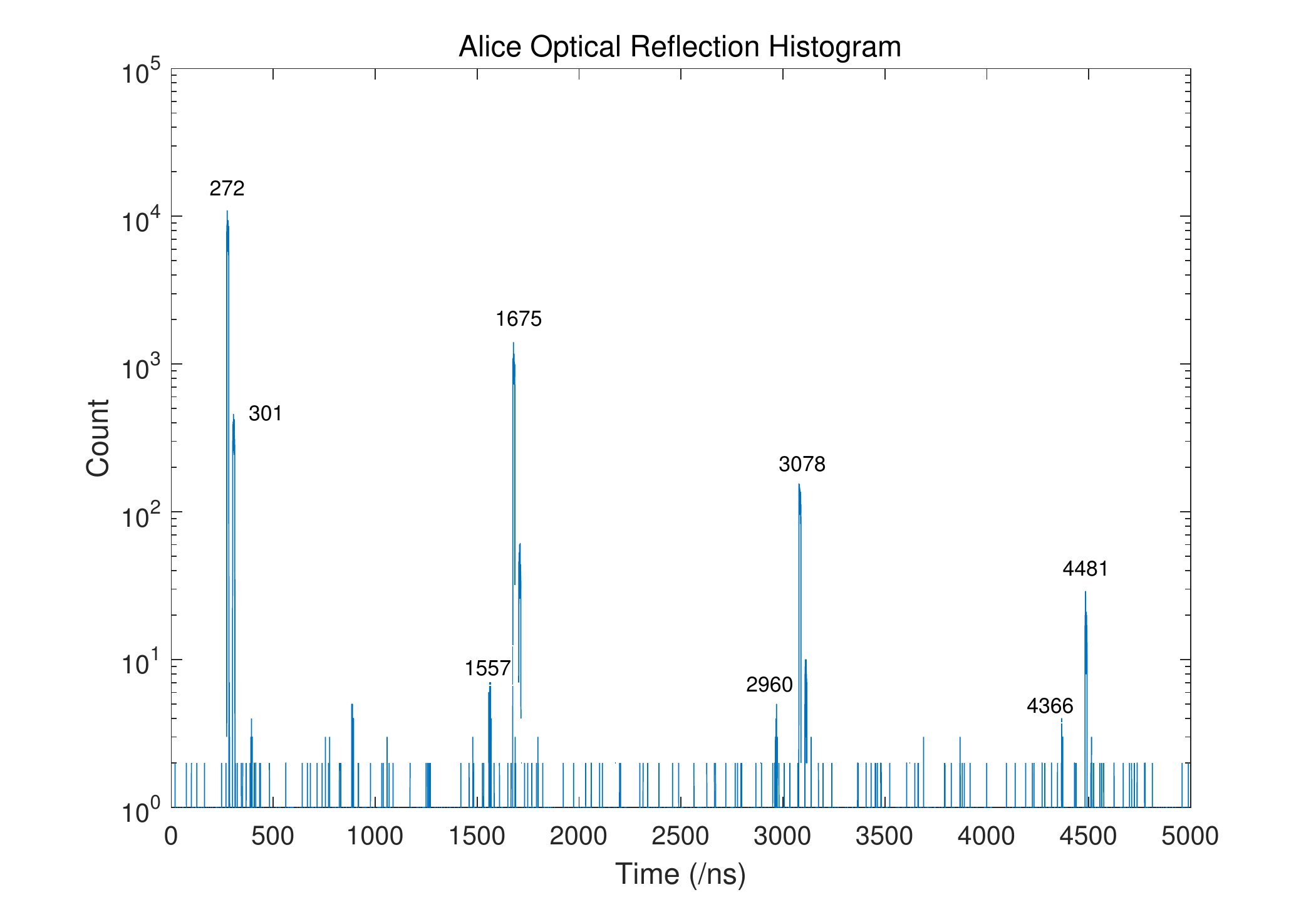}
    }
    \subfigure[]{
      \includegraphics[width=11cm]{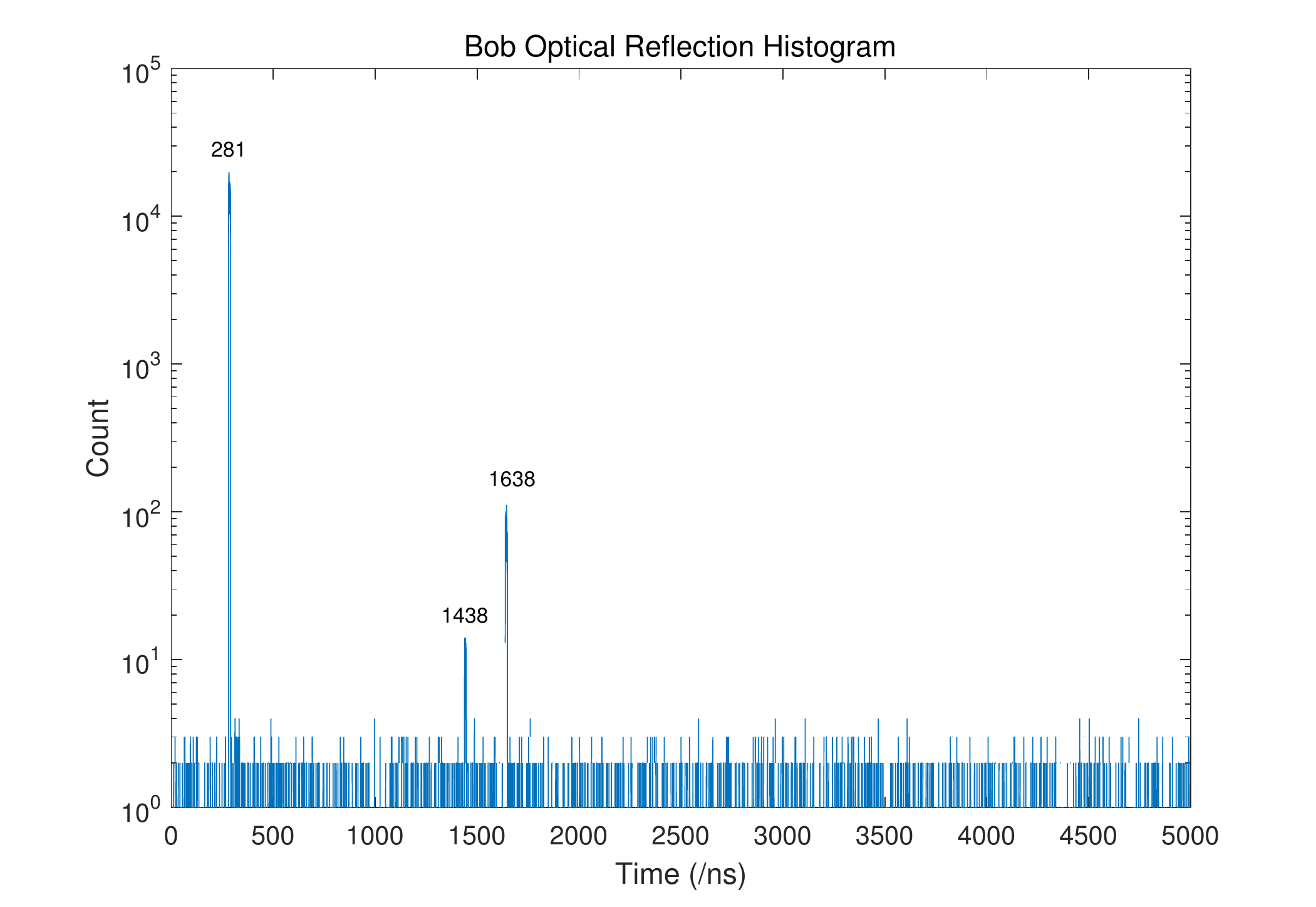}
    }
\caption{Optical reflection peaks for fiber length test. (a) Alice's optical reflection: the first detection occurs at 272 ns, the peak following (301 ns) is a noise peak due to false discrimination. The peaks at 1675 ns, 3078 ns, and 4481 ns are the reflections at SNSPD for the 1st, 2nd, and 3rd times. The peaks at 1557 ns, 2960 ns, and 4366 ns are reflection from the fiber-free space output at the pockels cell. The pockels cell is about 50 cm away from the fiber output port. (b) Bob's optical reflection: the first detection occurs at 281 ns, the peaks at 1638 ns is the reflection at SNSPD through the fiber link. The peak at 1438 ns is the reflection from the fiber-free space output at the pockels cell. The pockels cell is about 50 cm away from the fiber output port. Note: the polarization of the light incident onto the SNSPD is adjusted to maximize the reflection from SNSPD, while the polarization is adjusted to minimize the reflection in the Bell test experiment.}
\label{Fig:OTDR}
\end{figure}

In the experiment, we measure the fiber length by measuring the reflection: As shown in Fig.~\ref{Fig:OTDR}, The single photon signal arrives at the SNSPD, generates an electronic response with high possibility; with small possibility, the photon is reflected by the SNSPD chip, traveling to the source, get polarization rotated in the Sagnac loop, and traveling back to the SNSPD. This photon-travel doubles the distance from the source to the SNSPD, creating the second peak. We calculate the interval between the two peaks so we can calculate the fiber distance between the source and SNSPD. Because SNSPD is polarization sensitive, for the delay measurement, we maximize the reflection of light from SNSPD; but for the real experimental measurement, we minimize the light reflection. This is accomplished by changing the polarization of light. We measure the electronic cable length using a ruler.

\begin{figure}[htbp]
\centering
    \subfigure[]{
      \includegraphics[width=11cm]{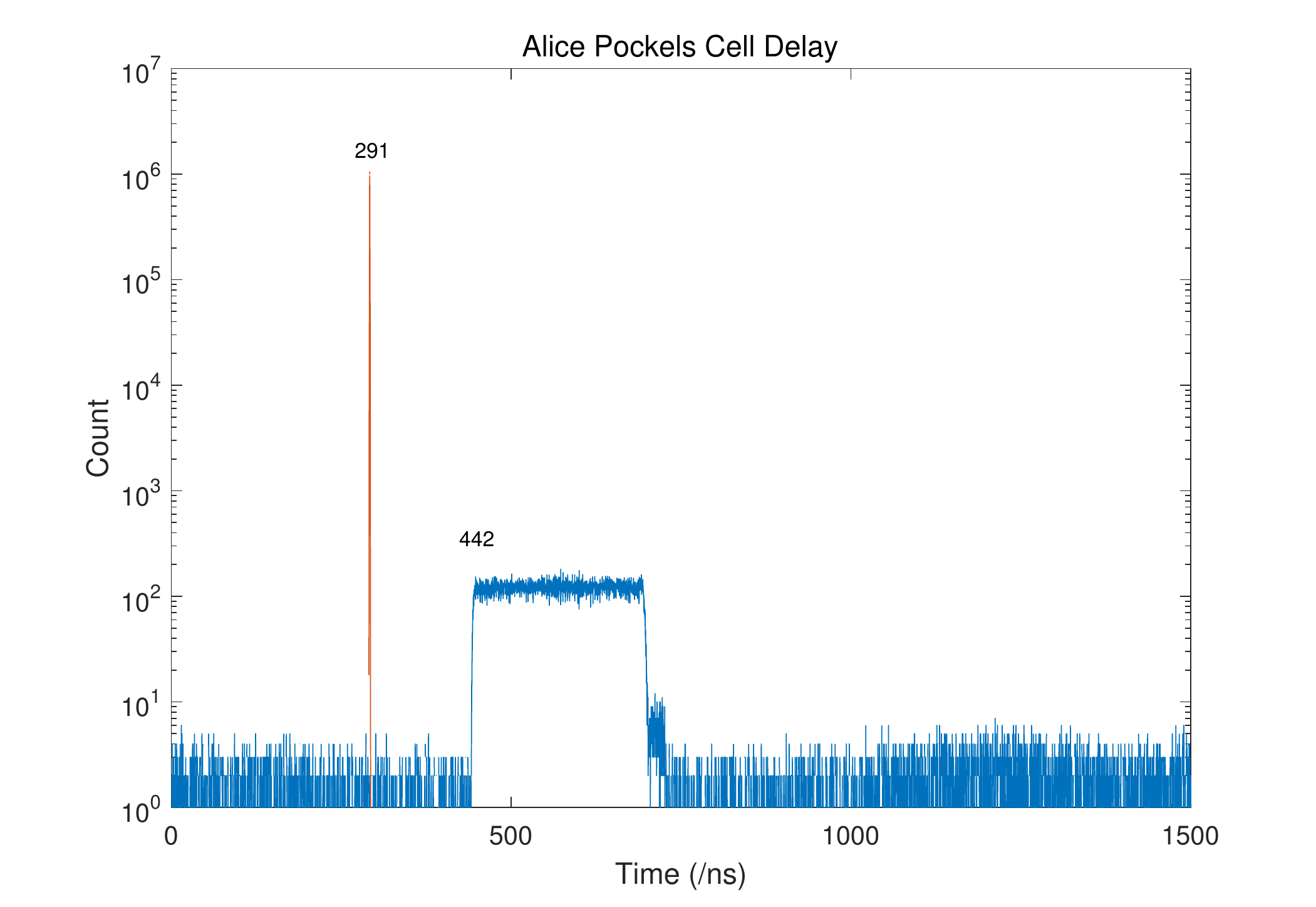}
    }
    \subfigure[]{
      \includegraphics[width=11cm]{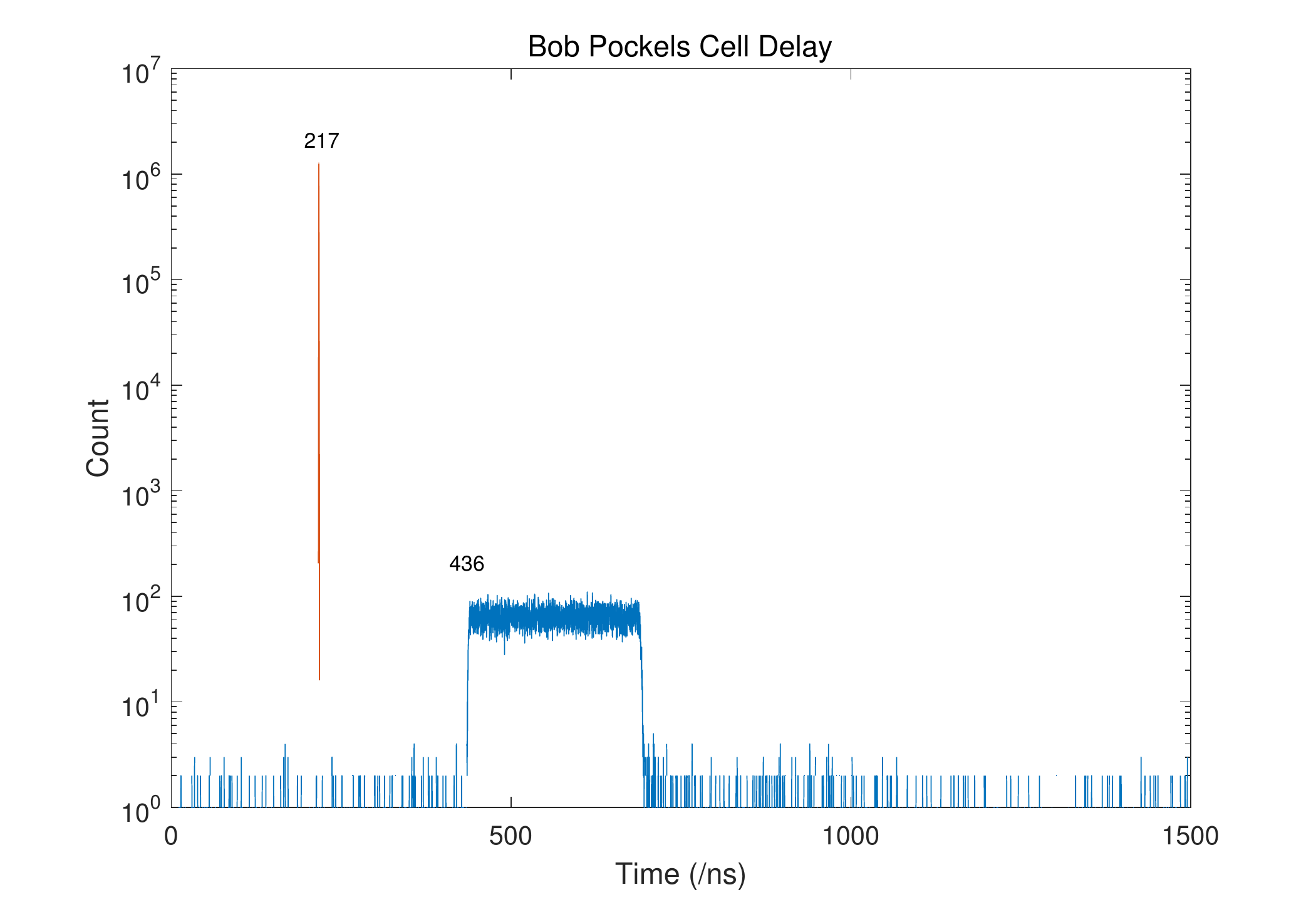}
    }
\caption{Delay test for Pockels cell. (a) Alice's Pockels cell delay: The Pockcels cell receives a trigger at 291 ns and the detection edge is at 442 ns. We measure Alice's cable length to be 4 meters, and discriminator delay to be 10 ns. (b) Bob's Pockels cell delay: The trigger is at 217 ns and the detection edge is at 436 ns. We measure Bob's cable length to be 12 meters, and discriminator delay to be 10 ns. }
\label{Fig:OTPC}
\end{figure}

To measure the time response of the Pockels cell's high voltage driver, we send a continuous-wave laser with horizontal polarization to the Pockels cell, and measure in the vertical basis using the SNSPD. The light will be blocked when there is no voltage applied, and will pass the measurement system when a half-wave voltage is applied. We apply a trigger signal to the high voltage driver, the interval between the trigger and the detection edge indicates the total delay of the Pockels cell modulation system, As shown in Fig.~\ref{Fig:OTPC}. By subtracting the time the signal travels in the fiber and the cable, and the delay caused by the discriminator, we calculate the effective fiber length between the pockels cell and the SNSPD chip. The measured fiber length, cable length, discriminating time, and the calculated Pockels cell driver response time are summarized in Tab.~\ref{Tab:FiberDist}.

\begin{table}[htbp]
\centering
  \caption{The fiber distances between Source and Measurement.}
\begin{tabular}{c|cccc}
\hline
 & Source-PC & PC-SNSPD & SNSPD-TDC & Pockels Cell Response\\
\hline
Alice & 132 m & 11 m & 4 m  & 62 ns\\
Bob   & 119 m & 20 m & 12 m & 50 ns\\
\hline
\end{tabular}
\label{Tab:FiberDist}
\end{table}

\subsection{Optimize mean photon number for optimum CHSH game value}

\begin{table}[htb]
\centering
  \caption{The Eberhard violation with different mean photon numbers.}
\begin{tabular}{c|c}
\hline
Mean photon number & CHSH violation\\
\hline
0.011 & $6.47\times10^{-5}$ \\
0.026 & $1.38\times10^{-4}$ \\
0.049 & $2.29\times10^{-4}$ \\
0.061 & $2.46\times10^{-4}$ \\
0.070 & $2.17\times10^{-4}$ \\
0.072 & $2.89\times10^{-4}$ \\
0.073 & $2.98\times10^{-4}$ \\
0.074 & $2.66\times10^{-4}$ \\
0.082 & $2.80\times10^{-4}$ \\
0.083 & $2.58\times10^{-4}$ \\
0.085 & $2.85\times10^{-4}$ \\
0.098 & $3.39\times10^{-4}$ \\
0.108 & $3.44\times10^{-4}$ \\
0.113 & $3.49\times10^{-4}$ \\
0.124 & $3.22\times10^{-4}$ \\
0.132 & $2.96\times10^{-4}$ \\
0.139 & $2.74\times10^{-4}$ \\
0.153 & $2.71\times10^{-4}$ \\
0.162 & $2.60\times10^{-4}$ \\
\hline
\end{tabular}
\label{tab:vioSum}
\end{table}

\begin{figure}[tbh]
\centering
     \resizebox{13cm}{!}{\includegraphics{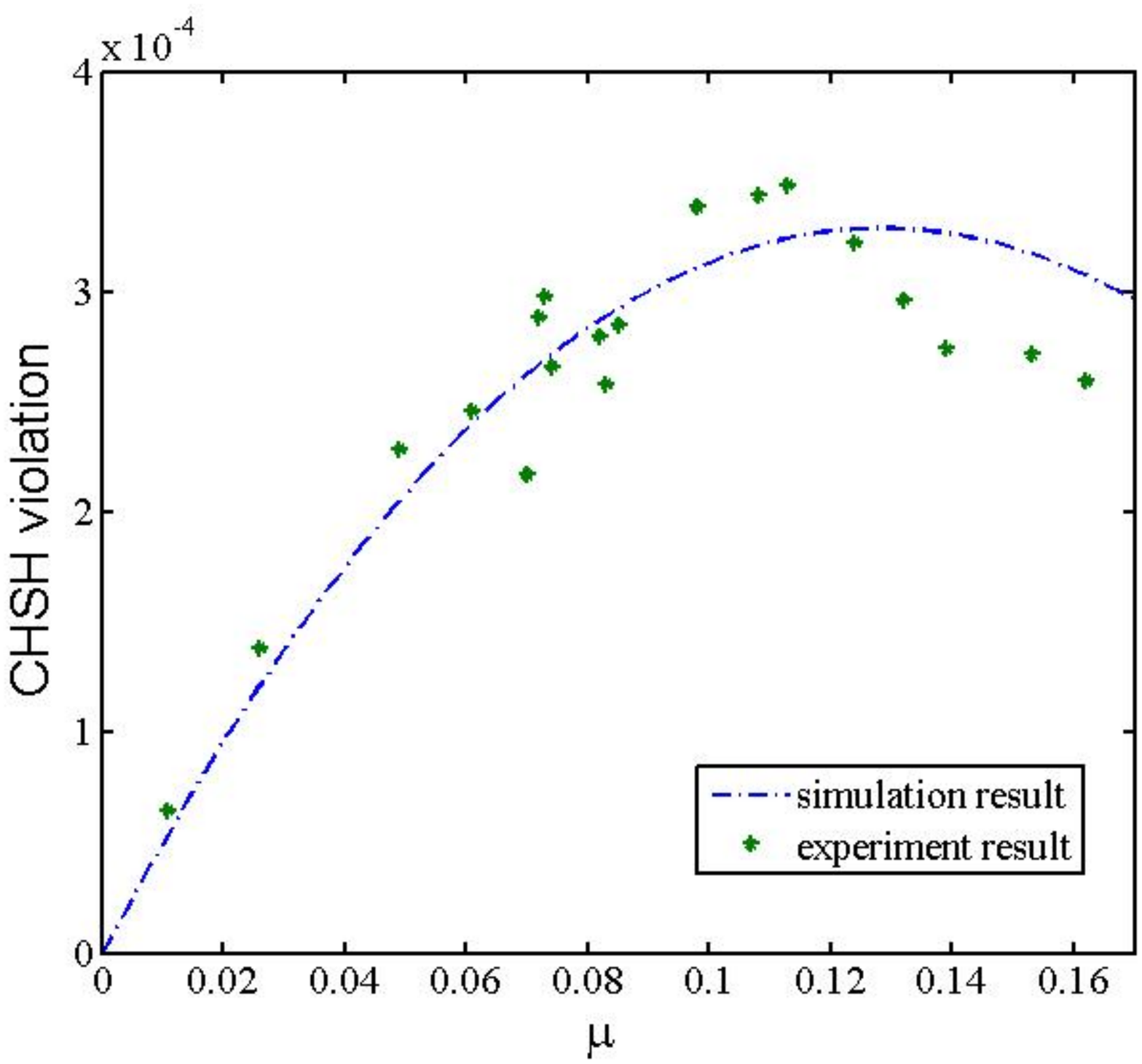}}
     \caption{Simulation/Experimental result for the relation between Bell inequality violation and mean photon number. In our simulation, we set dark count $p_B = 2\times10^{-5}$ and misalignment error $p_M = 5\times10^{-4}$.}
\label{Fig:randombbits}
\end{figure}

There is a optimized mean photon number to violate the CHSH game maximumly. Intuitively, the quantum states in most of the experimental trials are vacuum when the mean photon number is small. Increasing mean photon number will increase the proportion of non vacuum states and increase the Bell violation. The violation will decrease when the multi-photon effect becomes significant. We test the violation with experimental test in Tab.~\ref{tab:vioSum}, and perform a numerical simulation to optimize the mean photon number in below, the result is shown in Fig.~\ref{Fig:randombbits}. In the experiment, we conservatively select a sub-optimum intensity $\mu=0.07$.

Consider an SPDC source with a Poisson distribution,
\begin{equation}\label{}
  P(n) = \frac{\mu^n}{n!}e^{-\mu}
\end{equation}
where $\mu$ is the mean photon number. We denote $P(i)$ to be the $i$ pair photons case and denote ${J}_{n=i}$ to be Bell value for each case.
In the following, we simulate the Bell value with $0,1,2,3$ pairs of photon cases. We choose the threshold detector in the simulation and consider at most 3 pairs of photons.

For each pair of photons, we have nine results as shown in Table~\ref{Table:twopair}.
\begin{table}[hbt]
\centering
\caption{Possible events for single photon pair. }
\begin{tabular}{c|ccccccccc}
  \hline
  parties&1 &2 & 3 & 4 &5&6&7&8&9\\
  \hline
  Alice~and~Bob& 0,0& 0,1 & 0,u & 1,0 & 1,1 & 1,u & u,u & u,u & u,u\\
\hline
\end{tabular}\label{Table:twopair}
\end{table}
There are totally $9\times 9\times9=729$ exclusive events. When double click event happens, we randomly assign the output value to be 0, 1, or u, with probability $q_0$,  $q_1$, and $q_u$, respectively. We denote $q_0^a$ ($q_0^b$), $q_1^a$ ($q_1^b$) and $q_u^a$ ($q_u^b$), to be assignment probability for Alice's (Bob's) sides.

For a given input setting $x$ and $y$, we denote $p_i(x, y)$ to be the probability distribution for 9 events. In the end, we focus on the 0,0 case and denote $p^{n=j}(x, y)$ to be its probability generated by a $j$-pair photon case, $j$ can be $0,1,2$ or $3$ here. The simulation of the cases $j=0,1,2$ has already been studied in the previous work (Supplemental Materials of Ref.~\cite{PhysRevLett.120.010503}). In order to make the simulation better, we also consider the three-pair photon case here. We can calculate the probability $p^{n=3}(x, y)$ of obtaining 00 with three pairs of photons after the assignment by
\begin{equation}\label{eq:twopairab}
  p^{n=3}(x, y) = \sum_{ijk} \beta_{ijk} p_i(x, y)p_j(x, y)p_k(x, y),
\end{equation}
where the coefficients $\beta_{ijk}$ are chosen according to a $9\times 9\times9$ matrix. For the single party detections, the method is similar. We also take the misalignment error and dark count into consideration and we can calculate the final Bell value
\begin{equation}\label{}
  J \approx {J}_B+ P(1){J}_{n=1} + P(2){J}_{n=2}+P(3){J}_{n=3}.
\end{equation}
where ${J}_B$ is the normalized contribution from dark count

\subsection{System robustness}
\begin{figure}[tbh]
\centering
     \resizebox{13cm}{!}{\includegraphics{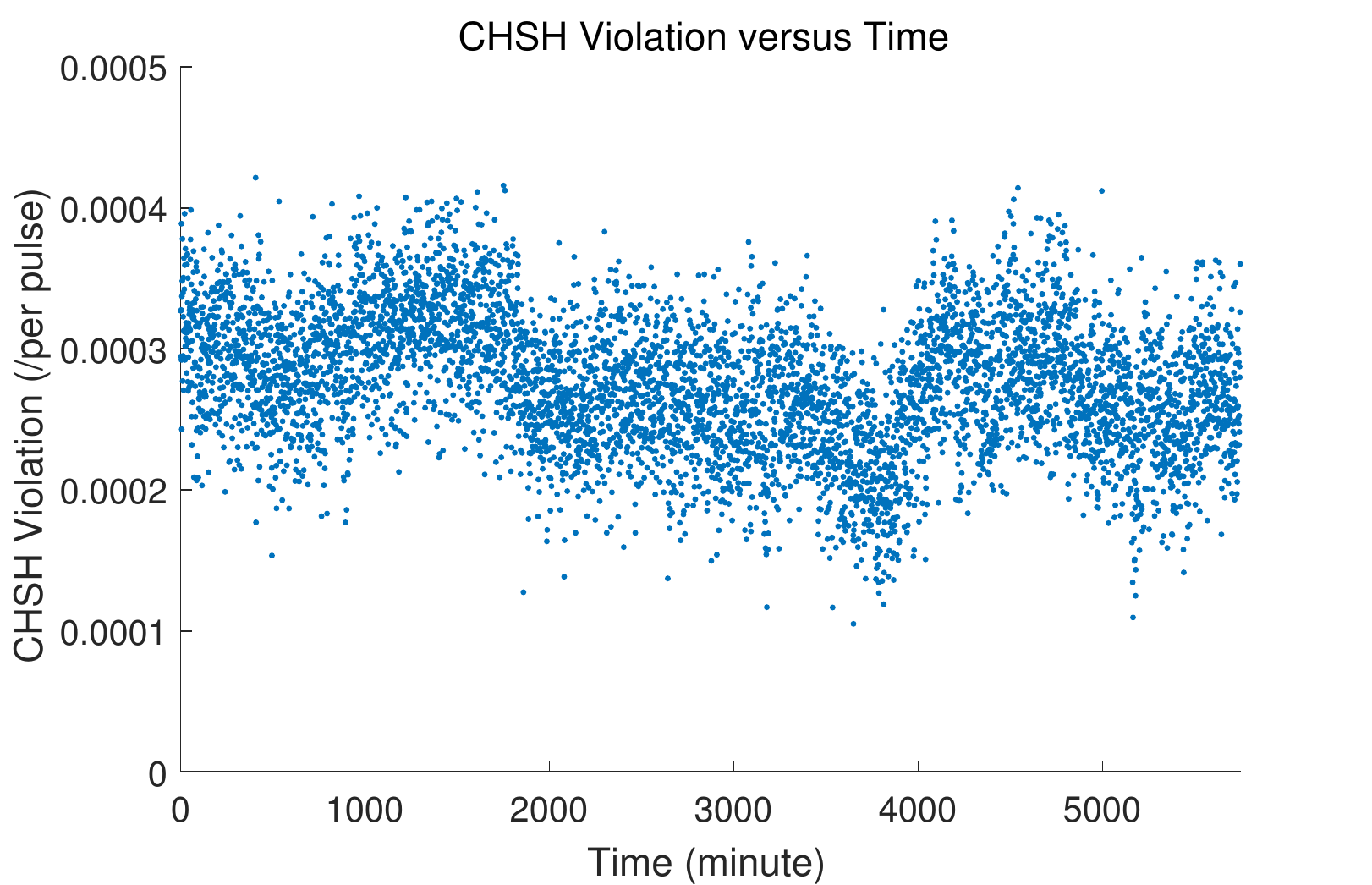}}
     \caption{The CHSH violation versus time.}
\label{Fig:chshtime}
\end{figure}

Fig.~\ref{Fig:chshtime} shows the measured CHSH violation value versus time. Every 60 seconds, we estimate the CHSH violation with the accumulated data. In the experiment, the optical alignment degrades slowly. We occasionally manually tweak the mirrors to restore the alignment. The CHSH violation remains at a sufficient significance level, which allows us to collect data continuously.

\subsection{Randomness extraction}
A Toeplitz extractor is used for randomness extraction from raw data \cite{Impagliazzo:Leftover:1989, frauchiger2013true, Liu_High_2018}. In total,$1.3790\times10^{11}$ bits of raw data are collected. A Toeplitz matrix with dimensions $m \times n =(6.2469\times10^7) \times (1.3790\times10^{11}$) is used to extract the final random numbers with a dimension $m=6.2469\times10^7$.

We used a blocked speed up algorithm with fast Fourier transform (FFT) in Toeplitz matrix multiplication \cite{Liu_High_2018}. The blocked FFT acceleration is not the optimism as the standard FFT algorithm, but it saves memory. The extraction is done on a personal computer with 16 Gbytes memory, with the original data  divided into 500 blocks for FFT acceleration. The whole calculation takes 11 hrs including data loading and computation.

\section{Experimental Results}
We complete $n=6.895\times10^{10}$ experimental trials in 95.77 experimental hours. The recorded experimental data are listed in Table \ref{tab:Eberhard}. The $J$-value for $n$-experimental trials as given by $J_n = J_{X_0 Y_0}+J_{X_0 Y_1}+J_{X_1 Y_0}+J_{X_1 Y_1}-3/4$, with $	J_{X_k Y_l} = (N_{ab=00|X_k Y_l}+N_{ab=11|X_k Y_l})/N_{X_k Y_l}$ when $(k,l)\in\{(0,0),(0,1),(1,0)\}$, and  $	J_{X_k Y_l} = (N_{ab=10|X_k Y_l}+N_{ab=01|X_k Y_l})/N_{X_k Y_l}$ when $(k,l)\in\{(1,1)\}$, is computed to be $\bar{J}=2.757 \times 10^{-4}$. $\bar{J}=2.757 \times 10^{-4}$ indicates that our CHSH game rejects local hidden variable models at a sufficient significance level (see Sect. III D for details).

By setting the expected CHSH game value to the one measured in the experiment, $\oexp = 2.757\times10^{-4}$, $\varepsilon_s = \varepsilon_{\ea} = 1 \times 10^{-5}$ and $\dest = \sqrt{10/n}= 1.2042 \times 10^{-5}$ and after applying an 137.90 Gb $\times$ 62.469 Mb Toeplitz matrix hashing, we obtain $6.2469 \times 10^{7}$ genuinely quantum-certified random bits, or 181.20 bits/s, with uniformity within $ 10^{-5}$. The stream of random bits pass the NIST statistic test suite. 

\begin{table}
\centering
  \caption{Recorded number of two-photon detection events for four sets of polarization state measurement bases $X_0 Y_0$, $X_0 Y_1$, $X_1 Y_0$ and $X_1 Y_1$ for $n=6.895\times10^{10}$ experimental trials. $a = 0$ or $1$ indicates that Alice detects a photon or not, the same $b$ for Bob. Mean photon number $\mu=0.07$, violation $J_n=2.757\times10^{-4}$.}
\begin{tabular}{ccccc}
\hline
Basis settings & $ab=00$ & $ab=10$ & $ab=01$ & $ab=11$\\
\hline
$X_0 Y_0$ & 17014507270 	& 58589512 	& 52352062 	& 112090418 \\
$X_0 Y_1$ & 16852014228 	& 217902589 	& 42594266 	& 121844486 \\
$X_1 Y_0$ & 16862026671 	& 46395448 	& 208761003 	& 124337061 \\
$X_1 Y_1$ & 16579373011 	& 326221778 	& 319412762 	& 13577435 \\
\hline
\end{tabular}
\label{tab:Eberhard}
\end{table}

\subsection{Statistical analysis of output randomness}
We obtain $6.2469\times10^7$ random bits. We use 62.469 Mbits data with section length set to 1.041 Mbits for NIST statistical test \cite{NIST_Tests}. As shown in Tab.~\ref{tab:nisttest}, the random bits successfully pass the NIST tests.

\begin{table}[htb]
\centering
  \caption{ Results of the NIST test suite using 62.469 Mbit of data (60 sequences of 1.041 Mbit) with the generated random numbers. }
\begin{tabular}{c|ccc}
\hline
Statistical tests & P value & Proportion & Result\\
\hline
Frequency				& 0.17828 & 1.000 & Success \\
BlockFrequency			& 0.73992 & 0.983 & Success \\
CumulativeSums			& 0.25360 & 1.000 & Success \\
Runs					& 0.13469 & 1.000 & Success \\
LongestRun				& 0.67178 & 1.000 & Success \\
Rank					& 0.04872 & 1.000 & Success \\
FFT						& 0.77276 & 0.967 & Success \\
NonOverlappingTemplate	& 0.08440 & 0.990 & Success \\
OverlappingTemplate		& 0.63712 & 1.000 & Success \\
Universal				& 0.96430 & 0.983 & Success \\
ApproximateEntropy		& 0.37814 & 0.983 & Success \\
RandomExcursions 		& 0.22430 & 0.990 & Success \\
RandomExcursionsVariant	& 0.50920 & 0.991 & Success \\
Serial 					& 0.10250 & 1.000 & Success \\
LinearComplexity		& 0.13469 & 1.000 & Success \\
\hline
\end{tabular}
\label{tab:nisttest}
\end{table}

\subsection{Test of no signaling}
Denote Alice's and Bob's random settings at each trial by $X$ and $Y$ with possible values $x, y\in \{0,1\}$. The joint-setting probability distribution is $\{p_{xy},x,y=0,1\}$, which is assumed to be fixed and known before running the test.  The measurement outcomes of Alice and Bob at each trial are denoted by $A$ and $B$ with possible values $a,b\in \{0,1\}$. Suppose that the experimentally observed frequency distribution is $\mathbf{f} \equiv \{p_{xy}f(ab|xy), a,b,x,y=0,1\}$. We would like to find out the no-signaling distribution $\mathbf{p}^{*}_{\text{NS}} \equiv \{p_{xy}p^{*}_{\text{NS}}(ab|xy), a,b,x,y=0,1\}$ that is optimally consistent with the observed frequencies. Here, we define the optimal distribution by minimizing a `distance' from the observed frequency distribution $\mathbf{f}$ to the set $\mathit{P}_\text{NS}$ of all the distributions $\mathbf{p}_{\text{NS}}$ satisfying the no-signaling principle. Particularly, we measure the distance from the observed frequency distribution $\mathbf{f}$ to a no-signaling probability distribution $\mathbf{p}_{\text{NS}}$ by the Kullback-Leibler (KL) divergence~\cite{Kullback1951}
\begin{equation}
D_{\text{KL}}(\mathbf{f}\parallel \mathbf{p}_{\text{NS}})=\sum_{a,b,x,y} p_{xy}f(ab|xy)\log_{2}\left(\frac{f(ab|xy)}{p_{\text{NS}}(ab|xy)}\right). \label{eq:KL_divergence}
\end{equation}
Hence, the optimal no-signaling distribution $\mathbf{p}^{*}_{\text{NS}}$
is the solution of the optimization
\begin{equation}
\min_{\mathbf{p}_{\text{NS}}\in \mathit{P}_\text{NS}} D_{\text{KL}}(\mathbf{f}\parallel \mathbf{p}_{\text{NS}}).
\label{eq:statistical_strength}
\end{equation}
Since the KL divergence $D_{\text{KL}}(\mathbf{f}\parallel \mathbf{p}_{\text{NS}})$ is a strictly convex function of the distribution $\mathbf{p}_{\text{NS}}$ and the domain $\mathit{P}_\text{NS}$ of the optimization problem in Eq.~\eqref{eq:statistical_strength} is a convex polytope~\cite{Barrett2005}, the solution $\mathbf{p}^{*}_{\text{NS}}$ of the above optimization is unique, which can be found by the sequential quadratic programming or the expectation-maximization algorithm as studied in Ref.~\cite{Zhang2011}.

Once the optimal no-signaling distribution $\mathbf{p}^{*}_{\text{NS}}$ is found, according to the method in Ref.~\cite{Zhang2011} any no-signaling distribution $\mathbf{p}_{\text{NS}}\in \mathit{P}_\text{NS}$ satisfies the following inequality
\begin{equation}
\sum_{a,b,x,y} \frac{f(ab|xy)}{p^{*}_{\text{NS}}(ab|xy)} p_{xy}p_{\text{NS}}(ab|xy) \leq 1.
\label{eq:pbr_inequality}
\end{equation}
The test statistics $R(ABXY)\equiv\frac{f(AB|XY)}{p^{*}_{\text{NS}}(AB|XY)}$ provide a way to perform a hypothesis test of the no-signaling principle~\cite{Zhang2011}. To perform the hypothesis test without assuming the trial results are i.i.d., before the $i$'th trial we need to construct the test statistics $R_i(A_iB_iX_iY_i)$ for this trial. For this purpose, we need to replace the experimentally observed frequency distribution $\mathbf{f}$ in Eqs.~\eqref{eq:KL_divergence},~\eqref{eq:statistical_strength} and~\eqref{eq:pbr_inequality} by a frequency distribution $\mathbf{f}_i$ estimated \emph{before} the $i$'th trial. The frequency distribution $\mathbf{f}_i$ can be estimated using all the trial results before the $i$'th trial or using only the most recent trial results in history.   Hence, the test statistics $R_i(A_iB_iX_iY_i)$ are called ``prediction-based ratios'', abbreviated as PRBs. Once the PBRs are constructed, after $n$ trials the $p$-value upper bound for rejecting the no-signaling principle is given by~\cite{Zhang2011}
\begin{equation}
p_{n}=\min\left(\left(\prod_{i=1}^{n}R_i(a_ib_ix_iy_i)\right)^{-1},1\right),
\end{equation}
where $a_ib_i$ and $x_iy_i$ are the measurement outcomes and setting choices at the $i$'th trial.

For the results observed in $95.77$ hours with $n= 6.895\times 10^{10}$, we process experimental data block by block, where each data block has 24,000,000 trials. We use the same PBRs for all the trials in a data block. This is allowed by the method in Ref.~\cite{Zhang2011}, as long as the PBRs for each data block indexed by $k$ are constructed and fixed before processing the data in this block. For the first data block, we use the trivial PBRs, i.e., $R(abxy)=1$ for all $a, b, x, y$ in this block. For a latter data block $k$ with $k>1$, considering the possible drift of experimental parameters over time, we construct the PBRs for the trial results in this block using the frequency distribution observed in the previous data block $(k-1)$.  Assuming that both the null and the alternative hypotheses have uniform settings, i.e., $p_{x_i,y_i}=1/4$ for all trials $i$ and all $x_i, y_i\in \{0,1\}$, we obtained the $p$-value upper bound $p_n=1$, suggesting no evidence of anomalous signaling in the experiment.

The method of PBRs is developed particularly for a hypothesis test where the experimental data provide a strong evidence against the null hypothesis. For the test of no signaling, if the no-signaling violation by experimental data is not strong, it is possible that the computed $p$-value upper bound with PBRs is not tight. We also checked whether our experimental data are in agreement with the no-signaling principle by a traditional hypothesis test where the i.i.d. assumption is required. In the experiment, there are four no-signaling conditions: the distribution of Alice's outcomes under the setting $x=0$ or $1$ is independent of Bob's setting choices, and the distribution of Bob's outcomes under the setting $y=0$ or $1$ is independent of Alice's setting choices. We performed a hypothesis test of each no-signaling condition with the two-proportion $Z$-test. We found the $p$-values of $0.139842$, $0.045396$, $0.474135$, and $0.226216$, which also suggest no evidence of violating the no-signaling principle by our experimental data.

\subsection{Test of local realism}
After obtaining the optimal no-signaling distribution $\mathbf{p}^{*}_{\text{NS}}$, we can find the optimal local realistic distribution $\mathbf{p}^{*}_{\text{LR}}$ according to
\begin{equation}
\mathbf{p}^{*}_{\text{LR}}=\text{argmin}_{\mathbf{p}_{\text{LR}}\in \mathit{P}_\text{LR}} D_{\text{KL}}(\mathbf{p}^{*}_{\text{NS}}\parallel \mathbf{p}_{\text{LR}}),
\label{eq:statistical_strength_lr}
\end{equation}
where $\mathit{P}_\text{LR}$ is the set of local realistic distributions.  As shown in Ref.~\cite{Zhang2011}, any local realistic distribution $\mathbf{p}_{\text{LR}}\in \mathit{P}_\text{LR}$ satisfies the following inequality
\begin{equation}
\sum_{a,b,x,y} \frac{{p}^{*}_{\text{NS}}(ab|xy)}{p^{*}_{\text{LR}}(ab|xy)} p_{xy}p_{\text{LR}}(ab|xy) \leq 1.
\label{eq:pbr_inequality_lr}
\end{equation}
Following the same procedure as the above for testing the no-signaling principle with PBRs, we can perform the hypothesis test of local realism without the i.i.d. assumption. The results observed in $95.77$ hours with $n= 6.895\times 10^{10}$ show that the $p$-value for rejecting the local realism after the experiment is upper bounded by $10^{-204792}$. This shows an extremely strong evidence against local realism provided by our experimental data, under the assumption that both the null and the alternative hypotheses have the uniform setting distribution at each trial.
\emph{If necessary}, we can also relax the assumption that the setting distribution is fixed and known by the strategy used in Ref.~\cite{Knill2017} for certifying randomness.

In conclusion, by the same analysis method and with the same assumption, our experimental data suggest no evidence against the no-signaling principle as well as an extremely strong evidence against local realism.

\bibliography{BibDIQRNG}
\end{document}